\documentclass[fleqn,usenatbib]{mnras}

\usepackage{newtxtext,newtxmath}

\usepackage[T1]{fontenc}

\DeclareRobustCommand{\VAN}[3]{#2}
\let\VANthebibliography\thebibliography
\def\thebibliography{\DeclareRobustCommand{\VAN}[3]{##3}\VANthebibliography}

\usepackage{subfig}
\usepackage{graphicx}	
\usepackage{amsmath}	
\usepackage{algorithm2e}
\usepackage{multicol}
\usepackage[para]{threeparttable}
\usepackage{xr}

\DeclareMathOperator*{\argmin}{arg\,min}
\newcommand{\norm}[1]{\left\lVert#1\right\rVert}

\DeclareMathOperator{\sinc}{sinc}

\title[CS-ROMER for Faraday depth reconstruction]{CS-ROMER: A novel compressed sensing framework for Faraday depth reconstruction}

\author[M. C{\'a}rcamo et al.]{
Miguel C{\'a}rcamo,$^{1,2,3}$\thanks{E-mail: miguel.carcamo@manchester.ac.uk (MC)}
Anna M.~M.~Scaife,$^{1,4}$
Emma L. Alexander$^{1}$
and J.~Patrick Leahy$^{1}$
\\
$^{1}$Jodrell Bank Centre for Astrophysics, Department of Physics and Astronomy, University of Manchester, Manchester, UK\\
$^{2}$University of Santiago of Chile (USACH), Faculty of Engineering, Computer Engineering Department, Chile\\
$^{3}$Center for Interdisciplinary Research in Astrophysics and Space Exploration (CIRAS), Universidad de Santiago de Chile\\
$^{4}$The Alan Turing Institute, Euston Road, London, NW1 2DB, UK
}

\date{Accepted XXX. Received YYY; in original form ZZZ}

\pubyear{2022}

\begin{document}
\label{firstpage}
\pagerange{\pageref{firstpage}--\pageref{lastpage}}
\maketitle

\begin{abstract}
	The reconstruction of Faraday depth structure from incomplete spectral polarization radio measurements using the RM Synthesis technique is an under-constrained problem requiring additional regularisation. In this paper we present {\tt cs-romer}: a novel object-oriented compressed sensing framework to reconstruct Faraday depth signals from spectro-polarization radio data. Unlike previous compressed sensing applications, this framework is designed to work directly with data that are irregularly sampled in wavelength-squared space and to incorporate multiple forms of compressed sensing regularisation. We demonstrate the framework using simulated data for the VLA telescope under a variety of observing conditions, and we introduce a methodology for identifying the optimal basis function for reconstruction of these data, using an approach that can also be applied to datasets from other telescopes and over different frequency ranges. In this work we show that the delta basis function provides optimal reconstruction for VLA L-band data and we use this basis with observations of the low-mass galaxy cluster Abell\,1314 in order to reconstruct the Faraday depth of its constituent cluster galaxies. We use the {\tt cs-romer} framework to de-rotate the Galactic Faraday depth contribution directly from the wavelength-squared data and to handle the spectral behaviour of different radio sources in a direction-dependent manner. The results of this analysis show that individual galaxies within Abell\,1314 deviate from the behaviour expected for a Faraday-thin screen such as the intra-cluster medium and instead suggest that the Faraday rotation exhibited by these galaxies is dominated by their local environments.

\end{abstract}




\section{Introduction}

The advent of broadband polarisation measurements with modern radio telescopes have enabled the widespread use of the RM Synthesis technique \citep{burn,brentjens}. RM Synthesis considers the Fourier relationship between polarized intensity (corrected by the spectral dependency) as a function of wavelength-squared ($P(\lambda^2)$ and the Faraday dispersion function (FDF). Thus, we can recover the polarized intensity as a function of Faraday depth, $\phi$, such that

\begin{equation}
	F(\phi) = \int_{-\infty}^{\infty}{\frac{P(\lambda^2)}{s(\lambda^2)} {\rm e}^{-2i\phi \lambda^2}~{\rm d}\lambda^2 },
	\label{eq:faraday-eq}
\end{equation}

where $s(\lambda^2)$ is the spectral dependence, where
\begin{equation}
	\label{eq:spectral-idx}
	s(\lambda^2) = \frac{I(\lambda^2)}{I(\lambda^2_0)} = \biggl(\frac{\lambda^2}{\lambda^2_0}\biggr)^{-\alpha/2}.
\end{equation}

Finally,
\begin{equation}
	\label{eq:p-pol}
	P(\lambda^2) = |P(\lambda^2)|{\rm e}^{2i\chi(\lambda^2)} = Q(\lambda^2) + iU(\lambda^2).
\end{equation}
Note that $P(\lambda^2)$ can be recovered from the polarized intensity as a function of Faraday depth such that
\begin{equation}
	\label{eq:p-fromfaraday}
	\frac{P(\lambda^2)}{s(\lambda^2)} = \frac{1}{\pi} \int_{-\infty}^{\infty}{F(\phi){\rm e}^{2i\phi \lambda^2}~{\rm d}\phi}.
\end{equation}

Faraday depth studies have led to a number of improvements in our understanding of astrophysical systems, including our own Galaxy \citep[e.g.][]{iacobelli2013}, external galaxies \citep[e.g.][]{cantwell2020} and on larger scales the intra- and inter-cluster medium \citep[e.g.][]{stuardi2021}. However, while the RM Synthesis technique is conceptually simple, in practical terms the implementation of this method and the interpretation of the resulting Faraday dispersion function are complicated by a number of different factors.

The first of these complications is that Faraday depth is the Fourier conjugate variable of $\lambda^2$, but radio spectrometers sample Stokes parameters $Q$ and $U$ uniformly in frequency, not $\lambda^2$.
This introduces additional computational complexity when reconstructing signals in Faraday depth space, similar to that encountered when imaging data from radio interferometers, which are natively sampled in time and (temporal) frequency but not in angular frequency. A common solution in both cases is to re-sample the original data onto a regular grid using a convolutional kernel, which then allows the use of a Fast Fourier Transform (FFT). For RM Synthesis this approach has been implemented, for example, in the widely used {\sc pyrmsynth}\footnote{\url{https://github.com/mrbell/pyrmsynth}} package used for the LOFAR Two-metre Sky Survey \citep{lofarpyrmsynth}. However, even in the case when such gridding is used, the inherent non-linear mapping of frequency to $\lambda^2$ results in a non-uniform distribution of measurements in $\lambda^2$-space. This non-uniform sampling is equivalent to a multiplicative weighting function, $W(\lambda^2)$, which therefore results in the convolution of the true Faraday dispersion function with a transfer function commonly known as the rotation measure transfer function, RMTF$(\phi)$ \citep{brentjens}, or, alternatively, the Rotation Measure Spread Function (RMSF).

A further complication is that $P(\lambda^2)$ does not exist at $\lambda<0$. Unlike the sky brightness distribution in the imaging case, $F(\phi)$ is hardly ever a purely real quantity and therefore a lack of measurements at $\lambda^2 < 0$ represents a fundamental limitation in attempting to reconstruct an unknown Faraday signal.

The effect of this is that attempts to deconvolve the RMTF from the Faraday depth spectra \cite[e.g.][]{heald2009} are inherently under-constrained due to the  $\lambda<0$ problem, which causes ambiguity between different solutions. Such under-constrained deconvolution processes may therefore introduce spurious structures, whilst also being unlikely to reconstruct all true physical structures \citep{JP2012,pratley2020}. Attempts have been made to address such under-constrained problems in the literature through the use of a suitable prior, or regularisation, during optimisation in order to compensate for the missing information \citep[e.g.][]{Akiyama_2017a, Cooray_2020, ndiritu2021, pratley_johnston-hollitt_gaensler_2021}.

The finite bandwidth of radio telescopes and hence the finite range of wavelength-squared, $\Delta(\lambda^2)$ available in the data is also a limitation, as is the potentially incomplete sampling over this bandwidth due to corrupting effects that require flagging or excision of specific frequency channels from an observation. Such effects include radio frequency interference (RFI) and instrumental problems, which typically result in approximately $20 - 40$\,\% of the frequency channels in a dataset being flagged \citep{Mauch_2020}. The effect of this additional excision adds to the problem of irregular sampling in $\lambda^2$-space and causes the form of the RMTF to worsen as the data sampling becomes increasingly affected.

In this work we present {\tt cs-romer}\footnote{\url{https://www.github.com/miguelcarcamov/csromer}}. A novel framework for reconstructing Faraday depth signals from noisy and incomplete spectro-polarimetric radio datasets. This framework is based on a compressed-sensing approach that addresses a number of outstanding issues in Faraday depth reconstruction in a systematic and scaleable manner. The structure of the paper is as follows: in Section~\ref{sec:cs} we introduce the technique of compressed sensing, the conditions under which it is beneficial and the optimization techniques used to apply the technique; in Section~\ref{sec:app_cs_ra} we give an overview of previous applications of compressed sensing in radio astronomy, including both interferometric imaging and Faraday depth reconstruction; in Section~\ref{sec:csromer} we present the {\tt cs-romer} compressed sensing Faraday depth reconstruction framework introduced in this work and in Section~\ref{sec:simdata} we demonstrate its application to simulated radio data, including evaluation and discussion of performance evaluation metrics on corrupted data; in Section~\ref{sec:realdata} we describe the use of real radio data from the Karl G. Jansky Very Large Array (VLA) towards the galaxy cluster Abell\,1314 and apply the {\tt cs-romer} framework to these data to extract magnetic field properties from this system; Finally, in Section~\ref{sec:conclusion} we draw our conclusions. The Supplementary material of this manuscript consists of Appendix~\ref{app:datacorruption}, which shows how evaluation metrics behave for simulated data with different amounts of removal and noise fractions, and Appendix~\ref{app:xray-xmm-analysis}, which presents the X-ray data reduction details and analysis.

\section{Compressed Sensing}
\label{sec:cs}

Compressed sensing (CS), also known as compressive sensing or sparse sampling, is a novel paradigm that contradicts the Nyquist-Shannon theorem \citep{nyquist, shannon} for data acquisition under certain circumstances. CS states that sparse high-dimensional signals subject to a specific constraint can be recovered from far fewer samples or measurements than used by traditional methods, which dictate sampling at a rate at least twice the signal bandwidth. In the application to signal processing, the method is formulated in a way to obtain an accurate signal from an incompletely sampled dataset. The degree of success of this method depends on the amount of information that can be supplied to constrain the signal solution while being consistent with the measurements \citep{thompsonbook_2017}. These constraints are sparsity, non-negativity, compactness, and the smoothness of the signal.

For example, let $x$ be a measured signal of length $N$, or a representation of it in some basis. This can be written as $x=\sum_i^N c_i\phi_i$, where $\{\phi_i\}_{i=1}^N$ is a set of linearly independent vectors and $\{c_i\}_{i=1}^N$ are unique coefficients that represent signal $x$. If we let $\Phi$ be the $N\times N$ matrix with columns given by $\phi_i$ we can represent this relationship in a compact manner such that $x=\Phi c$.

Technically, $x$ is not measured directly. Instead, $M < N$ linear measurements are acquired using a $M \times N$ \textit{sensing matrix} or a linear transform. This can be mathematically represented as $y = Ax$. Here, matrix $A$ is considered a \textit{dimensionality reduction} since it maps $\mathbb{R}^N$, where $N$ is generally large, into $\mathbb{R}^M$, where $M$ is much smaller than $N$. In this scenario (when $M \ll N$ measurements are taken) an ill-posed problem arises: multiple signals $x$ result in the same measurements $y$, posing the question as to how the original signal can be recovered from measurements.

To fully recover a signal $x$, a compressible representation as a $k$-sparse vector must be imposed. Typically, signals are not themselves sparse, but they can have a sparse representation in some known basis or dictionary $\Phi$. Mathematically, a signal is defined to be sparse or compressible if contains a small number, $k$, of non-zero or significant values. . In this case we can refer to $x$ being $k$-sparse considering $x=\Phi c$ and its $L_0$ norm or number of non-zero elements as $\norm{c}_0 \leq k$. Some examples of commonly used bases/dictionaries are wavelets, curvelets, Fourier components, etc. The most well known wavelet families are the Haar wavelet family \citep{Haar1910} and the Daubechies wavelet family \citep{daubechies1992ten}.

There are a wide variety of methods available to reconstruct a signal from measured data. Given measurements, $y$, and knowing that the signal, $x$, is sparse we can solve the following optimization problem
\begin{equation}
	\hat{x} = \argmin_{x} \norm{x}_0 \; \text{subject to} \; x\in \mathcal{B}(y),
\end{equation}

where $\mathcal{B}(y)$ ensures that $\hat{x}$ is consistent with the data, $y$. In an ideal case $\mathcal{B}(y) = \{x: Ax = y\}$ if the  measurements are not corrupted by noise. Otherwise, $\mathcal{B}(y) = \{x: \norm{y - Ax}_2^2 \leq \epsilon\}$, where $\epsilon$ is governed by the degree of noise present in the data. The main problem with this approach is that the zero-norm function, which counts the non-zero elements of a vector, is non-convex and finding a solution that approximates the true minimum is generally a non-deterministic polynomial-time hard (NP-hard) problem and computationally very intensive \citep{np-hard}.

\citet{donoho_cs} and \citet{candestao2016} discovered that under general conditions it is possible to solve (relax) the above problem by using basis pursuit or enforcing convexity in the $\norm{\cdot}_0$. That is, considering a Laplacian prior or a $L_1$-regularization as
\begin{equation}
	\hat{x} = \argmin_{x} \norm{x}_1 \; \text{subject to} \; x\in \mathcal{B}(y),
	\label{eq:$L_1$min}
\end{equation}
and that $\mathcal{B}(y)$ is convex, then the problem has been translated into a computationally feasible one that can be solved with efficient methods from convex optimization. Here, the $L_1$ norm of a vector is defined as the sum of the absolute values of each component of the vector:
\begin{equation}
	\norm{x}_1 = \sum_{t=1}^N |x_t|.
	\label{eq:$L_1$norm}
\end{equation}

In the presence of noise, Equation~\ref{eq:$L_1$min} can also be written as an unconstrained problem:
\begin{equation}
	\hat{x} = \argmin_{x} \frac{1}{2}\norm{Ax-y}_2^2 + \eta \norm{x}_1\; ,
	\label{eq:$L_1$min_2}
\end{equation}
where the first term is equivalent to the squared sum of the data residuals and $\eta$ is a regularization parameter that determines the relative importance between minimizing the $L_1$-norm and the measurement residuals.

\subsection{RIP, Sparsity and Incoherence}
\label{sec:rip}

To ensure sparse recovery of a signal in presence of noise, CS relies on two conditions: sparsity and incoherence, applied through the Restricted Isometry Property (RIP).

At the beginning of Section \ref{sec:cs} we defined our signal as $x=\Phi c$, where $c$ is considered a $k$-sparse vector since it contains a small number of $k$ non-zero elements.

On the other hand, coherence is defined as the largest correlation between any row of a sensing matrix $A$ and any column of a representation basis $\Phi$. The less the coherence between these two matrices, the fewer measurements are needed to recover the signal \citep{candesieee}. A common example of maximal incoherence occurs when using a Fourier sensing matrix and a canonical (delta functions) representation basis.

If both sparsity and incoherence conditions are satisfied, then it can be said that matrix $A\Phi$ meets the $k$-RIP, and therefore, all subsets of $k$ columns taken from $A\Phi$ are nearly orthogonal. Thus, the optimization problem on Equation \ref{eq:$L_1$min_2} can be solved with high probability.

Detailed definitions and mathematical proofs of these conditions can be found in \cite{candesieee} and \cite{Foucart:2013:MIC:2526243}.

\subsection{Regularization}
\label{sec:regularization}

Until here we have defined the problem using a regularization for sparsity such as $L_1$-norm. However in CS, Total Variation (TV), often computed as
\begin{equation}
	\text{TV}(x)= \sum_{i}|x_{i+1} - x_i|\;,
	\label{eq:tv1d}
\end{equation}
in its 1-dimensional version and
\begin{equation}
	\text{TV}(x)= \sum_{i,j}|x_{i+1,j} - x_{i,j}| + |x_{i,j+1} - x_{i,j}|\;,
	\label{eq:tv2d}
\end{equation}
in its 2-dimensional version, and Total Squared Variation (TSV) computed as
\begin{equation}
	\text{TSV}(x) = \sum_i |x_{i+1} - x_{i}|^2 \;,
	\label{eq:tsv1d}
\end{equation}
in its 1D version and
\begin{equation}
	\text{TSV}(x) = \sum_{i,j} |x_{i+1,j} - x_{i,j}|^2 + |x_{i,j+1} - x_{i,j}|^2 \;,
	\label{eq:tsv2d}
\end{equation}
in its 2D version have been also used in the literature (see \citet{akiyama2018faraday, eht-blackhole}). Note that while TV can also be seen as the $L_1$-norm for adjacent pixel differences and minimizes the gradients, thereby favoring smoother signals \citep{thompsonbook_2017} with flat regions separated by edges \citep{eht-blackhole}, TSV only favors those signals with smooth edges.

Another well-known regularization is the $L_2$-norm, that can be seen as the negative logarithm of a Gaussian prior distribution. However, this prior will fail when producing a sparse solution, whereas if an $L_1$-norm is used the signal can be exactly recovered. In turn, a traditional linear method based on $L_2$-minimization fails in reconstructing the original signal. This is because the Laplacian distribution is highly peaked and bears heavy tails in comparison to the Gaussian distribution \citep{2009MNRAS.395.1733W}. In other words, the Laplacian prior has a greater distribution mass around zero, while the Gaussian distribution is more diffuse around its center.

\subsection{CS Optimization methods}
\label{sec:optimization}

$L_1$-minimization can be solved using general-purpose convex optimization methods such as steepest descent, conjugate gradient (CG) and Quasi-Newton Methods \citep{nocedal2006numerical, numerical_r, elad}. Additionally, proximal splitting methods such as \emph{Alternating-direction method of multipliers} \citep[ADMM;][]{admm_1, admm_2} and \emph{Simultaneous-direction method of multipliers}  \citep[SDMM;][]{2014MNRAS.439.3591C} have been used to optimize a particular objective functions. In general, proximal methods solve optimization problems of the form
\begin{equation}
	\min_{x \in \mathbb{R}^N} f_1(x) + ... + f_s(x),
\end{equation}
where $f_1(x), ..., f_s(x)$ are convex lower semi-continuous functions from $\mathbb{R}^N$ to $\mathbb{R}$, that are not necessarily differentiable.

Iterative algorithms have been developed to solve the $L_1$-minization problem. These methods estimate coefficients of the sparse signal iteratively until they reach a convergence criterion. The most common algorithms are the \emph{Orthogonal Matching Persuit} \citep[OMP;][]{OMP} and \emph{Iterative Hard Thresholding} \citep[IHT;][]{IHT}. On each iteration, the OMP method correlates the columns of $A$ with the signal residual (obtained by subtracting the estimation of the signal with the observed measurement) until reaching a limit on the number of iterations or accomplishing the requirement  $y \approx A\hat{x}$. On the other hand, IHT starts from an initial signal estimate, $\hat{x}_0 = 0$, and iterates a gradient descent step using the hard thresholding operator on $x$, which sets all entries to zero except for the entries of $x$ with largest magnitude. These steps are repeated until meeting a convergence criterion.

\section{Applications of Compressive Sensing to radio astronomy}
\label{sec:app_cs_ra}

CS is a technique that suits big computing/data problems since it compresses data sizes substantially using basis coefficients, a compression that can be considered an additional constraint to the ill-posed problem and can result in a better regularization. The main applications of this theory to image synthesis in radio astronomy were made by \citet{2009MNRAS.395.1733W}, \citet{Li}, \citet{2012MNRAS.426.1223C, 2014MNRAS.439.3591C}, and \citet{purify2}.

\subsection{Applications of Compressive Sensing to 1D data in Radio Astronomy}
\label{sec:app1d}

Even though $P(\lambda^2) = Q(\lambda^2)+iU(\lambda^2)$ can be considered to be a complex image cube, one can take each individual line of sight (pixel) in $P$ and optimize it to recover a Faraday rotation measurement. As a result, we expect a set of optimized Faraday Spectrum vectors, $F$. In other words, another image cube. Therefore, most RM Synthesis papers consider reconstruction methods using one dimensional data. Applications of CS to this problem were made by \cite{frick, li20102, Andrecut_2011, akiyama2018faraday, craft2}.

There are two things to note in these works that reconstruct the Faraday depth function by using CS. They create data in Faraday depth space and then transform to $\lambda^2$-space using a Discrete Fourier Transform (DFT). However, doing this implies that the $\lambda^2$-space is implicitly regularly-spaced. It is worth noting that an irregular $\lambda^2$-space adds a layer of difficulty to the reconstruction problem. We also note that in these works the simulations do not account for frequency channel excision due to RFI, instrumental problems or calibration.

\section{CS-ROMER}
\label{sec:csromer}

{\tt CS-romer} is an object-oriented implementation of a CS-framework that solves the problem stated in Equation~\ref{eq:$L_1$min_2}, extending the objective function by adding TV and TSV regularizations in order to impose edge-smoothed constraints when needed. Consequently, we aim to solve the following optimization problem
\begin{equation}
	\hat{x} = \argmin_{x} \Biggl\{ f(x) + g(x) \Biggr\} \,
	\label{eq:gen-objf}
\end{equation}
where
\begin{equation}
	f(x) = \frac{1}{2} \sum_{i=1}^M \biggl|\frac{P^o_i - P^m_i(x)}{\sigma_i}\biggr|^2 =\frac{1}{2}\sum_i \biggl|\frac{r_i}{\sigma_i}\biggr|^2\;,
	\label{eq:gen-chi2}
\end{equation}
is commonly seen in the literature as a $\chi^2$ term, with $\{r_i\}$ the residuals, and is considered a convex smooth function; $g(x)$ is a convex function that is non-differentiable in some region and can be either
\begin{equation}
	\nonumber g(x) = \eta_1 \text{$L_1$}(x) + \eta_2 \text{TV}(x)\;,
	\label{eq:objf}
\end{equation}
or
\begin{equation}
	g(x) = \eta_1 \text{$L_1$}(x) + \eta_2 \text{TSV}(x)\;.
	\label{eq:objf}
\end{equation}
In these equations $x$ can be the Faraday depth signal itself or a wavelet representation of it; $L_1$, TV and TSV are defined in Equations~\ref{eq:$L_1$norm}, \ref{eq:tv1d} and \ref{eq:tsv1d}, respectively; $P^o$ is an $M$-size vector of the observed  polarized intensity measurements as a function of $\lambda^2$, and $P^m$ denotes the model measurements, which are functions of the signal estimate calculated using Equation~\ref{eq:p-fromfaraday}. Finally, $\sigma^2_k$ is the noise variance for channel $k$. Before continuing, it is important to highlight that our framework will first calculate characteristic parameters involved in Faraday depth reconstruction as defined in \cite{brentjens} and \cite{li20102}. First, the framework calculates $\lambda^2_{\text{min}}$ and $\lambda^2_{\text{max}}$. Then, $\Delta \lambda^2$ and $\delta \lambda^2$ are computed as
\begin{equation}
	\Delta \lambda^2 = \lambda^2_{\text{max}} - \lambda^2_{\text{min}}
\end{equation}
and
\begin{equation}
	\delta \lambda^2 = \langle \lambda^2_i - \lambda^2_{i-1}\rangle\;,
\end{equation}
respectively. These can in turn be used to define the
resolution in Faraday depth space, the largest scale in Faraday depth to which the data are inherently sensitive and the maximum observable Faraday depth. In \cite{brentjens} and our framework these are approximated as

\begin{equation}
	\delta \phi \approx \frac{2\sqrt{3}}{\Delta \lambda^2}\;,
\end{equation}
\begin{equation}
	\text{max-scale} \approx \frac{\pi}{\lambda^2_{\text{min}}}\;,
\end{equation}
and
\begin{equation}
	|\phi_{\text{max}}| \approx \frac{\sqrt{3}}{\delta \lambda^2}\;.
\end{equation}
These parameters can be also associated with the FWHM of the main peak of the RMTF, the maximum width that a recovered structure has and the maximum Faraday depth to which one has more than 50\% sensitivity, respectively. The framework selects the cell-size in Faraday depth space as $\phi_R = \delta \phi / \rho$, where $\rho$ is a selectable oversampling factor. Following \cite{li20102} we can use $|\phi_{\text{max}}|$ to calculate the length of the grid in Faraday depth space as
\begin{equation}
	n = \Biggl\lfloor \frac{2|\phi_{\text{max}}|}{\phi_R} \Biggr\rfloor.
\end{equation}

Since our observed measurements belong to an irregular space and Faraday depth space is on a regular grid the framework offers two options to estimate the model data. The first and most intuitive option is to grid the observed measurements, in which case the framework estimates $P^m$ from the Faraday spectra using a DFT or FFT. The second option is to use the Non-Uniform Fast Fourier Transform \citep[NUFFT;][]{jimaging4030051}, which approximates a Fourier transform on non-uniform sampled data using an FFT on the signal and then a min-max interpolation \citep{1166689}.

Having decided on a method with which to approximate the model data points, the next step is to use an optimization method to minimize Equation~\ref{eq:objf}. In this work we have chosen the Fast Iterative Shrinkage-Thresholding Algorithm \citep[FISTA;][]{fista}. This method is an enhancement of the ISTA method \citep{ista}. These methods are based on the proximal gradient method \citep{proximal}. To be concise, the scaled proximal operator is defined as
\begin{equation}
	\text{prox}_{\eta f}(y) = \argmin_x\biggl\{f(x) +\frac{1}{2\eta} ||x-y||_2^2\biggr\},
	\label{eq:proximal}
\end{equation}
and can be interpreted as a kind of gradient step for function $f$ with $\eta>0$ as its step size \citep{proximal}. A FISTA with constant step size and $k$ iterations is shown in Algorithm~\ref{alg:fista}.
\begin{algorithm}
	\caption{FISTA with constant step size}\label{alg:fista}
	\textbf{Input}: $\eta$, $x_0$ \\
	\textbf{Step 0.} $y_1 = x_0$, $t_1 = 1$\\
	\textbf{Step k.} $(k \geq 1)$ Compute\\
	\begin{align}
		x_k     & = \text{prox}_{\eta g}(y_k - \nabla f(y_k))                              \\
		t_{k+1} & = \frac{1+\sqrt{1+4t_k^2}}{2}                                            \\
		y_{k+1} & = x_k + \biggl(\frac{t_k - 1}{t_{k+1}}\biggl)\biggl(x_k - x_{k-1}\biggr)
	\end{align}
\end{algorithm}
The gradient calculation of $\nabla f(y_k)$ in Algorithm~\ref{alg:fista} is made by estimating the model visibilities as explained above. It is important to highlight that if we are using a basis representation we need to reconstruct the Faraday depth space from coefficients before estimating the model measurements. Then, we apply the Inverse Discrete Fourier Transform (IDFT) to the residuals as in Equation~\ref{eq:faraday-eq}.

To compute the proximal of our separable function $g(x) = g_1(x) + g_2(x)$ is to useful to recall that
\begin{eqnarray}
	\nonumber    \text{prox}_{\eta (g_1+g_2)} (y) &=& \text{prox}_{\eta_1 (g_1)} (\text{prox}_{\eta_2 (g_2)} (y))\\
	\nonumber     &=& \text{prox}_{\eta_2 (g_2)} (\text{prox}_{\eta_1 (g_1)} (y))\;.
\end{eqnarray}
as proved by \cite{pathwise}, particularly for a function composed of $L_1$-norm plus TV regularization. Later on in \cite{proximalmap}, this was proved more generally for any regularization pair. In some cases, the proximal operator of a function can be calculated analytically, such as the case of the $L_1$-norm \citep[see e.g.][]{Woodworth_2016}. Nevertheless, not all proximal operators can have their solutions derived analytically. Therefore, for TV and TSV regularizations we use the python package \texttt{proxtv}, which solves the proximal optimization problem using Fast Newton-type Methods \citep{proxtv1, proxtv2}.

Many attempts have been made to automatically calculate the regularization coefficients, $\eta_1$ and $\eta_2$, \citep[see e.g.][]{KARL2005183, Hansen00thel-curve, Belge_2002, SHI201872}. For this framework, and only for $L_1$ regularization, we adopt the error bound calculation in \cite{2014MNRAS.439.3591C, purify2}. Note that as the error bound decreases, the model will start to fit the noise and as it gets too large the model will not be able to properly fit signal structures. Therefore, the $\eta_1$ parameter in this framework is automatically calculated as
\begin{equation}
	\label{eq:eta}
	\eta_1 = \sqrt{(M + 2 \sqrt{M})}\sigma_{\phi} \;,
\end{equation}
where $M$ is the number of measurements in $\lambda^2$-space and $\sigma_{\phi}$ is defined in Table \ref{tab:noise}. Moreover, for observations when $M$ is large, this can be estimated as
\begin{equation}
	\eta_1 = \sqrt{M}\sigma_{\phi} \;.
\end{equation}

Additionally, we have adopted the adaptive regularization method from \cite{li20102}, where on each iteration $1\,\sigma_{\phi}$ is subtracted from $\eta_1$. This constrains reconstructed signals to have residuals with an rms of $1\,\sigma_{\phi}$. However, empirical examination using simulated data show that residuals obtained using the constraint coefficient given in Equation~\ref{eq:eta} and a Dirac-$\delta$ function basis for reconstruction differ from those made with any other wavelet transform by a factor of $\sim 2$. These results imply that the algorithm over-fits the noise by this factor when using wavelet transforms and a $1\,\sigma_{\phi}$ threshold. Consequently, the {\tt cs-romer} framework introduces an additional factor of 2 into Equation~\ref{eq:eta} to avoid such over-fitting when using wavelets other than the Dirac basis.

In the case where the signal can be represented by a wavelet dictionary, our framework offers the set of discrete wavelet transforms (DWT) from the package \texttt{pywt} \citep{Lee2019}. However, sometimes the DWT can cause problems due to its shift variance and poor directional properties \citep{uwt-denoising}. As an alternative, and as a way to reconstruct both thin and thick Faraday structures, we have added the Undecimated Wavelet Transform (UWT) 1D functions from \cite{Lee2019} to our framework. Another motivation for this addition is that it has been shown that the use of the thresholding function and UWT can improve results in image denoising and reconstruction applications \citep{starck, 10.1093/mnras/stx1547}, and even though there have been studies of wavelets in the RM Synthesis context, the UWT has not yet been investigated.

\subsection{Spectral dependency}
\label{sec:spectral}

As shown in Equation \ref{eq:faraday-eq}, the complex polarized intensity can be corrected by the spectral dependence of the total intensity as a function of $\lambda^2$. Although \cite{brentjens} explain the effect of the spectral index on Faraday depth spectra, there is no discussion about how resolution changes as function of wavelength-squared affect this process. For extended emission, such changes result in lower signal for some lines of sight at smaller values of $\lambda^2$, therefore affecting the spectral dependency. Here we consider three ways to deal with this problem:
\begin{enumerate}
	\item Use the same average spectral index for all lines of sight. According to \cite{brentjens} this will help to correct the spectra of those sources that are detected in individual channels maps and that only show up after averaging across the full band.
	\item Restore all channels in the spectral cube using the PSF corresponding to the reference frequency used in MFS reconstruction and use the spectral index map provided by the MFS reconstruction, e.g. using {\tt tclean}, to calculate $s(\lambda^2)$ (see Equation \ref{eq:spectral-idx}).
	\item Restore the spectral cube using the PSF corresponding to the lowest measured frequency channel and an appropriate spectral index, either the same across the cube as in point (i), or derived directly from the restored cube using individual lines of sight.
\end{enumerate}

The {\tt cs-romer} framework provides all three options to the user, who must specify their preference.

\subsection{De-rotation of Galactic RM}
\label{sec:derotgal}

To correct for foreground Galactic rotation, {\tt cs-romer} provides an option to read the HEALPix image model from \cite{faradaysky2020} and use a bilinear interpolation of those data to calculate the mean and standard deviation of the Galactic RM for each line of sight. We denote the interpolated mean image as $\phi_{\text{GAL}}$ and the de-rotation of this value can be applied directly as a shift in $\lambda^2$ space. Thus, {\tt cs-romer} optionally applies this operation directly to the $P(\lambda^2)$ cube as follows
\begin{eqnarray}
	\hat{P}(\lambda^2) &=& P(\lambda^2)e^{-2i\phi_{\text{GAL}}\lambda^2},
	\label{eq:gal-derot}
\end{eqnarray}
where $\hat{P}(\lambda^2)$ denotes the de-rotated complex polarization.

We note that the object-oriented implementation of {\tt cs-romer} (publicly available on \href{https://www.github.com/miguelcarcamov/csromer}{Github}) makes it easy for users to incorporate additional functionality that allows one to specify alternative foreground maps. This will be useful for future studies as improved Galactic foreground measures become available, or in cases where the user wishes to reproduce a specific foreground subtraction as a matter of preference.

\section{Application to Simulated Data}
\label{sec:simdata}

To illustrate and test the {\tt cs-romer} compressed sensing framework, we use the following models to simulate Faraday depth spectra analytically for three scenarios. We use a frequency range for all scenarios of $1.008-2.031$\,GHz, consistent with the real VLA data considered later in this work.

\begin{figure*}
	\includegraphics[width=\textwidth]{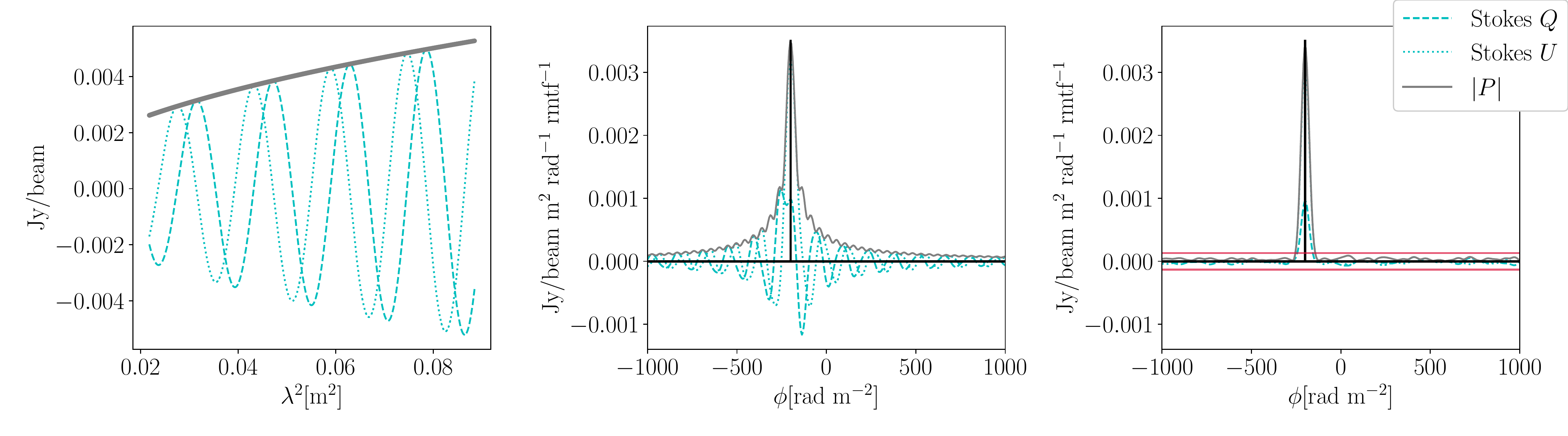}
	\caption{Scenario 1. On the left we show the simulated polarization data as a function of $\lambda^2$. In the center, the dirty Faraday depth spectrum of the same data. On the right, the reconstructed Faraday depth spectrum from a noisy and incomplete realisation of these data with $\sigma_{QU} =0.7$\,mJy/beam with a 30\% removal fraction, using the delta basis function. The theoretical $\pm5\sigma_{\phi}$ noise boundary for complex Faraday depth is shown as red lines. }
	\label{fig:fdthin}
\end{figure*}

First, we consider polarized emission from the lobe of a radio galaxy as:
\begin{equation}
	P_{\rm rg}(\lambda^2) =  S_{\nu_0}\left(\frac{\lambda^2}{\lambda_0^2}\right)^{-\alpha/2}\exp (2i\phi_1 \lambda^2)
\end{equation}
where $S_{\nu_0}$ is the intensity at the reference frequency, $\alpha$ is the spectral index and $\phi_1$ is the single Faraday depth value. This can be also seen as a thin Faraday source in Faraday-space. We set an intensity of $0.035$\,Jy/beam, a Faraday depth of $\phi_1 = -200$\,rad\,m$^{-2}$ and a spectral index $\alpha=1.0$ (see Figure~\ref{fig:fdthin}).

\begin{figure*}
	\centering

	\includegraphics[width=\textwidth]{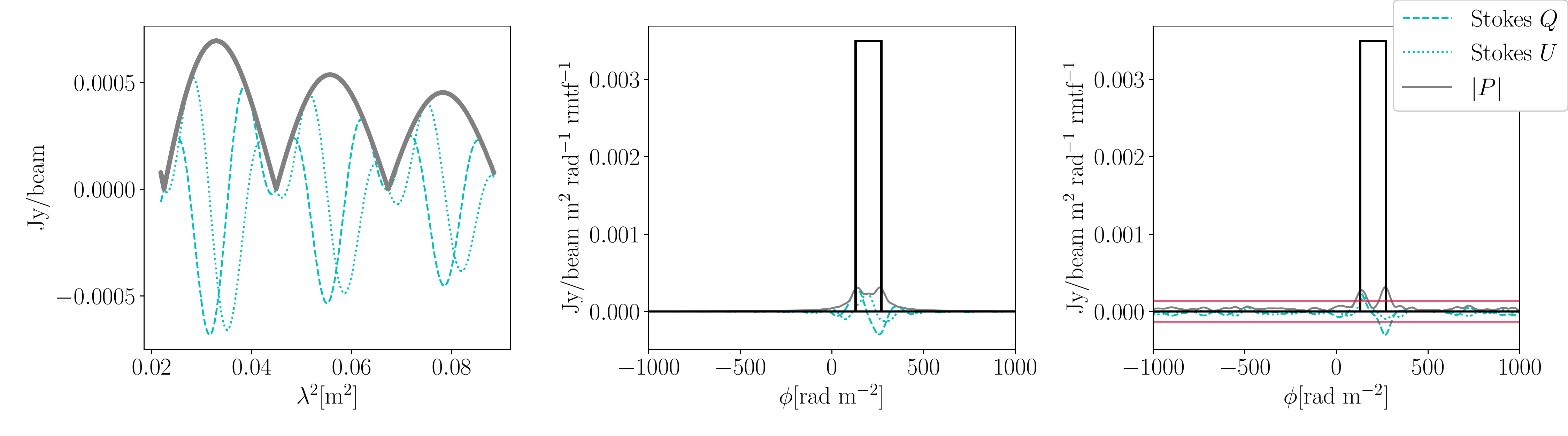}
	\caption{Scenario 2. On the left we show the simulated polarization data as a function of $\lambda^2$. In the center, the dirty Faraday depth spectrum of the same data. On the right, the reconstructed Faraday depth spectrum from a noisy and incomplete realisation of these data with $\sigma_{QU} =0.7$\,mJy/beam with a 30\% removal fraction, using the delta basis function. The theoretical $\pm5\sigma_{\phi}$ noise boundary for complex Faraday depth is shown as red lines.}
	\label{fig:fdthick}
\end{figure*}
\begin{figure*}
	\centering

	\includegraphics[width=\textwidth]{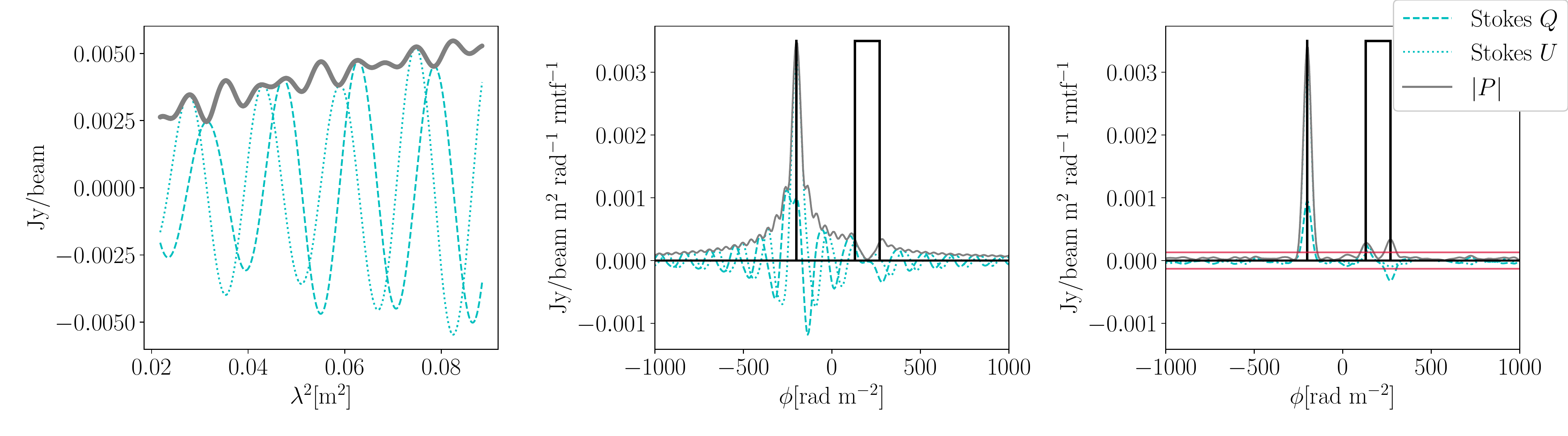}
	\caption{Scenario 3. On the left we show the simulated polarization data as a function of $\lambda^2$. In the center, the dirty Faraday depth spectrum of the same data. On the right, the reconstructed Faraday depth spectrum from a noisy and incomplete realisation of these data with $\sigma_{QU}=0.7$\,mJy/beam with a 30\% removal fraction, using the delta basis function. The theoretical $\pm5\sigma_{\phi}$ noise boundary for complex Faraday depth is shown as red lines.}
	\label{fig:fdthickthin}
\end{figure*}

The second scenario has been adopted from the top-hat function in \cite{brentjens} which represents a Faraday-thick component. We have generalised the equations from that work such that a component of width $\phi_{\rm fg}$ can be centered at any $\phi_0$ value:
\begin{equation}
	F_{\rm gal}(\phi) = \begin{cases}
		S_{\nu_0}(\phi_{\rm fg})^{-1} & \phi_0-\frac{\phi_{\rm fg}}{2}~<~\phi~<~\phi_0+\frac{\phi_{\rm fg}}{2} \\
		0                             & {\rm elsewhere}
	\end{cases}
\end{equation}
\begin{eqnarray}
	\nonumber P_{\rm gal}(\lambda^2) &=& \frac{S_{\nu_0}}{\lambda^2 \phi_{\rm fg}} \sin{\bigl(\lambda^2 \phi_{\rm fg}\bigr)} e^{\bigl(2i\lambda^2\phi_0\bigr)}\\
	&=& S_{\nu_0}e^{\bigl(2i\lambda^2\phi_0\bigr)}\sinc{\bigl(\lambda^2 \phi_{\rm fg}\bigr)}
\end{eqnarray}

In this case we set an intensity of $0.035$\,Jy/beam and a width of 140\,rad\,m$^{-2}$, centred on $\phi_0 = 200$\,rad\,m$^{-2}$. We choose this value because it is very close to the width of the maximum recoverable structure for the given frequency range (see Figure~\ref{fig:fdthick}).

Finally, the third scenario is a mixture of a Faraday thin source and a Faraday thick source. This is made by simply summing the resulting complex polarization intensities from previous scenarios:
\begin{equation}
	P_{\rm tot}(\lambda^2) = P_{\rm gal}(\lambda^2) + P_{\rm rg}(\lambda^2);
\end{equation}
see Figure~\ref{fig:fdthickthin}. The parameters for all three scenarios are summarised in Table~\ref{tab:JVLA-sim}.
\begin{table}
	\centering
	\caption{Abell\,1314 VLA simulation details.}
	\label{tab:JVLA-sim}
	\begin{tabular}{@{}ccccc@{}}
		\hline
		Scenario & $\phi_1$     & $\phi_{\text{fg}}$ & $\phi_0$     & $S_{\nu_0}$ \\
		& [rad m$^-2$] & [rad m$^-2$]       & [rad m$^-2$] & [Jy/beam]   \\\hline
		1        & -200         & -                  & -            & 0.0035      \\
		2        & -            & 140                & 200          & 0.0035      \\
		3        & -200         & 140                & 200          & 0.0035      \\\hline
	\end{tabular}
\end{table}

To be consistent with channel excision due to RFI and calibration we have randomly removed data samples from our simulations. In particular, for all three scenarios we have taken out different fractions of the total number of data points, i.e. frequency channels. Therefore, different data sets were made with removal fractions in a range from $0.0$ to $0.9$ using the method described in \cite{ndiritu2021}. Finally, to make the scenarios even more realistic we also add different levels of Gaussian random noise, with levels that scale as a fraction of the peak value from $0.0$ to $0.9$ in regular steps of $0.1$.

\subsection{Evaluation Metrics}
\label{sec:eval}

This section aims to explain the different evaluation metrics used in this work in order to evaluate the Faraday Rotation Measurement Synthesis framework and the different solutions obtained from the simulated scenarios described in Section~\ref{sec:simdata}.

\subsubsection{Peak-signal-to-noise ratio (PSNR)}
\label{sec:psnr}

Before continuing, it is important to define what is meant by the restored Faraday signal. Similarly to imaging in radio interferometry, most of the time the resulting \emph{model} signal will not represent the system's resolution. Therefore, to create the \emph{restored} signal we convolve the resulting model signal with a zero mean Gaussian that has a standard deviation equivalent to that of the RMTF.

The peak-signal-to-noise (PSNR) metric computes the maximum value of a signal divided by the corrupting noise that affects its fidelity. Since Faraday-space signals are complex, we take the maximum of the absolute value of the restored signal as the peak. The noise is calculated as the standard deviation of the residuals in Faraday depth space. This can be written as
\begin{equation}
	\text{PSNR} = \frac{\max{\{|F|\}}}{\sigma},
\end{equation}
where $F$ represents the restored Faraday depth signal and $\sigma$ is the standard deviation of the residuals in the same space, calculated as the arithmetic mean of the rms noise on the real and imaginary components of the Faraday depth spectrum:
\begin{equation}
	\sigma = \frac{\sigma_{\text{REAL}}+ \sigma_{\text{IMAG}}}{2}.
\end{equation}

\subsubsection{Root Mean Squared Error (RMSE)}
\label{sec:rmse}

One way to assess the reconstruction and how well the model fits the observed measurements is calculating the Root Mean Squared Error (RMSE). When the reconstruction reaches its end, the framework returns the residuals (c.f. \ref{eq:gen-chi2}) and we can use these to calculate the RMSE as
\begin{equation}
	\text{RMSE} = \sqrt{\frac{\sum_{i} |r_i|^2}{2n}}.
\end{equation}

Therefore, the RMSE can be interpreted as the average distance between the model and the observed data. The lower this distance is, the better the fit is.

\subsubsection{Sparsity}
\label{sec:sparsity}

As has been noted previously, sparsity can be defined as the number of non-zero elements, $k$, in a data vector of length $n$. The fewer the number of non-zero elements, the sparser the data. Here we calculate the percentage of non-zero elements, $s$, in the resulting parameters after optimization. This can be the sparse Faraday-space representation of the data, or the basis coefficient representation. When working with coefficients that have a real representation, this can be written as the proportion between non-zero coefficients and the length of the coefficient vector
\begin{equation}
	s = 100 \cdot k/n.
\end{equation}
However, since Faraday-space is complex, we need to calculate sparsity for the real and for the imaginary parts as
\begin{equation}
	s = 100 \cdot (k_{\text{REAL}} + k_{\text{IMAG}})/(2\cdot n).
\end{equation}

\subsubsection{Model selection criteria: AIC and BIC}
\label{sec:aic}

\begin{figure*}
	\centering
	\includegraphics[width=\textwidth]{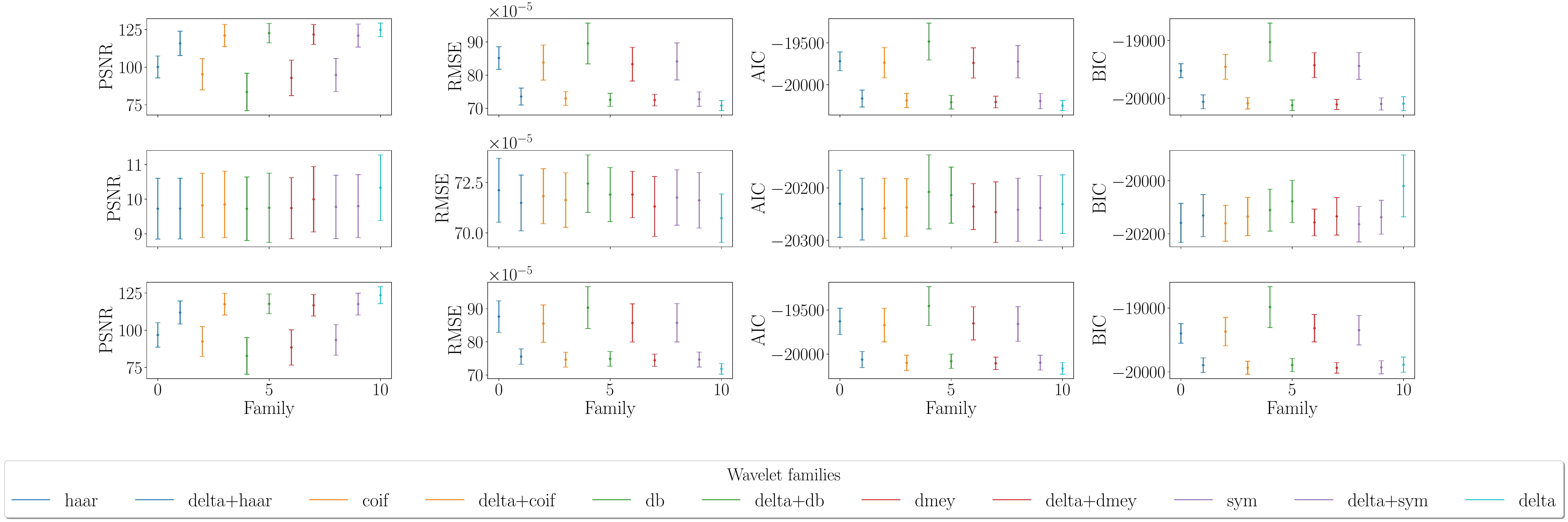}
	\caption{PSNR, RMSE, AIC and BIC means and standard deviations for each one of the families using all their discrete wavelet transforms. First row shows scenario 1, second row for scenario 2 and third row for scenario 3.}
	\label{fig:wavelet_study}
\end{figure*}

\begin{figure*}
	\centering
	\includegraphics[width=\textwidth]{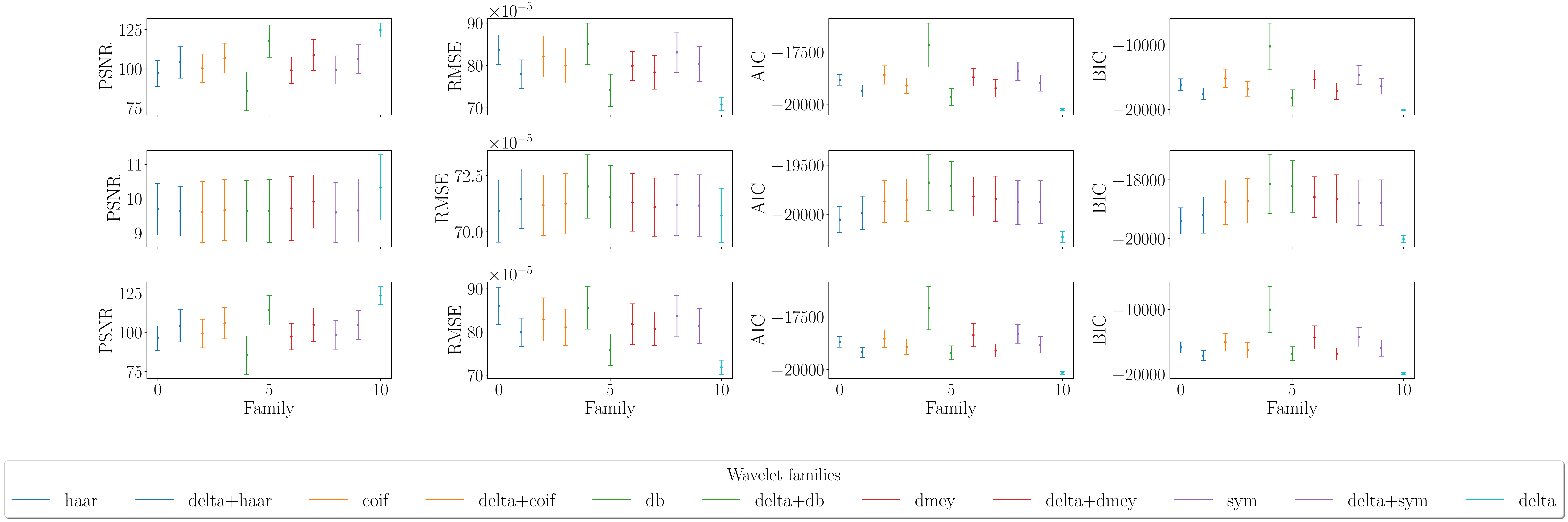}
	\caption{PSNR, RMSE, AIC and BIC means and standard deviations for each one of the families using all their undecimated wavelet transforms. First row shows scenario 1, second row for scenario 2 and third row for scenario 3.}
	\label{fig:undecimated_wavelet_study}
\end{figure*}

\begin{table*}
	\caption{AIC, BIC, PSNR and RMSE for delta function basis.}
	\label{tab:deltafunctionbasis}
	\begin{tabular}{@{}lccc@{}}
		\hline
		& \textbf{Scenario 1}              & \textbf{Scenario 2}              & \textbf{Scenario 3}              \\ \hline
		\textbf{AIC}  & $-20248.78 \pm 60.24$            & $-20231.11 \pm 55.93$            & $-20163.09 \pm 66.42$            \\
		\textbf{BIC}  & $-20093.34 \pm 120.40$           & $-20019.56 \pm 116.71$           & $-19887.35 \pm 119.22$           \\
		\textbf{PSNR} & $124.79 \pm 4.49$                & $10.33 \pm 0.95$                 & $123.53 \pm 5.72$                \\
		\textbf{RMSE} & $(70.83 \pm 1.52)\times 10^{-5}$ & $(70.73 \pm 1.21)\times 10^{-5}$ & $(71.85 \pm 1.62)\times 10^{-5}$ \\ \hline
	\end{tabular}%
\end{table*}

\begin{table*}
	\centering
	\caption{AIC, BIC, PSNR and RMSE for discrete wavelets transforms (WT) and undecimated wavelet transforms (UWT) with the minimum AIC.}
	\label{tab:best-family}
	\resizebox{\textwidth}{!}{%
		\begin{tabular}{@{\extracolsep{3pt}}lcccccc@{}}
			\hline
			& \multicolumn{2}{c}{\textbf{Scenario 1}} & \multicolumn{2}{c}{\textbf{Scenario 2}} & \multicolumn{2}{c}{\textbf{Scenario 3}}                                                                                                            \\ \hline
			& \multicolumn{1}{c}{\textbf{WT}}         & \multicolumn{1}{c}{\textbf{UWT}}        & \multicolumn{1}{c}{\textbf{WT}}         & \multicolumn{1}{c}{\textbf{UWT}}  & \multicolumn{1}{c}{\textbf{WT}}  & \multicolumn{1}{c}{\textbf{UWT}}  \\
			\cline{2-3} \cline{4-5} \cline{6-7}
			\textbf{Best wavelet family} & \texttt{dmey}                           & \texttt{haar}                           & \texttt{sym}                            & \texttt{haar}                     & \texttt{coif}                    & \texttt{haar}                     \\
			\textbf{AIC}                 & $-19739.53 \pm 180.53$                  & $-18823.76 \pm 260.78$                  & $-20241.77 \pm 60.25$                   & $-20054.28 \pm 133.15$            & $-19671.22 \pm 191.22$           & $-18690.92 \pm 252.91$            \\
			\textbf{BIC}                 & $-19427.18 \pm 216.16$                  & $-16156.66 \pm 893.47$                  & $-20163.74 \pm 66.63$                   & $-19398.75 \pm 446.16$            & $-19369.09 \pm 222.83$           & $-15865.54 \pm 861.84$            \\
			\textbf{PSNR}                & $92.86 \pm 11.78$                       & $97.11 \pm 8.30$                        & $9.78 \pm 0.92$                         & $9.70 \pm 0.75$                   & $92.44 \pm 10.03$                & $96.22 \pm 7.80$                  \\
			\textbf{RMSE}                & $(83.30 \pm 5.06)\times 10^{-5}$        & $(83.74 \pm 3.46)\times 10^{-5}$        & $(71.76 \pm 1.38) \times 10^{-5}$       & $(70.92 \pm 1.38) \times 10^{-5}$ & $(85.49 \pm 5.64)\times 10^{-5}$ & $(85.97 \pm 4.26) \times 10^{-5}$ \\ \hline
		\end{tabular}
	}
\end{table*}

\begin{table*}
	\centering
	\caption{AIC, BIC, PSNR and RMSE for delta basis function combined with discrete wavelets transforms (D+WT) and undecimated wavelet transforms (D+UWT) with the minimum AIC.}
	\label{tab:best-deltafamily}
	\resizebox{\textwidth}{!}{%
		\begin{tabular}{@{\extracolsep{3pt}}lcccccc@{}}
			\hline
			& \multicolumn{2}{c}{\textbf{Scenario 1}} & \multicolumn{2}{c}{\textbf{Scenario 2}} & \multicolumn{2}{c}{\textbf{Scenario 3}}                                                                                                               \\ \hline
			& \multicolumn{1}{c}{\textbf{D+WT}}       & \multicolumn{1}{c}{\textbf{D+UWT}}      & \multicolumn{1}{c}{\textbf{D+WT}}       & \multicolumn{1}{c}{\textbf{D+UWT}} & \multicolumn{1}{c}{\textbf{D+WT}} & \multicolumn{1}{c}{\textbf{D+UWT}} \\
			\cline{2-3} \cline{4-5} \cline{6-7}
			\textbf{Best wavelet family} & \texttt{db}                             & \texttt{db}                             & \texttt{dmey}                           & \texttt{haar}                      & \texttt{dmey}                     & \texttt{db}                        \\
			\textbf{AIC}                 & $-20207.76 \pm 81.40$                   & $-19644.59 \pm 414.10$                  & $-20246.30 \pm 57.76$                   & $-19984.59 \pm 168.32$             & $-20104.92 \pm 70.98$             & $-19211.01 \pm 332.61$             \\
			\textbf{BIC}                 & $-20120.36 \pm 90.66$                   & $-18232.65 \pm 1254.23$                 & $-20134.18 \pm 70.59$                   & $-19203.41 \pm 612.72$             & $-19937.21 \pm 82.66$             & $-16832.39 \pm 1064.69$            \\
			\textbf{PSNR}                & $122.61 \pm 6.49$                       & $117.66 \pm 10.28$                      & $10.00 \pm 0.95$                        & $9.64 \pm 0.72$                    & $116.75 \pm 7.20$                 & $114.13 \pm 9.43$                  \\
			\textbf{RMSE}                & $(72.55 \pm 1.96)\times 10^{-5}$        & $(74.14 \pm 3.78)\times 10^{-5}$        & $(71.31 \pm 1.49) \times 10^{-5}$       & $(71.47 \pm 1.32) \times 10^{-5}$  & $(74.45 \pm 1.85)\times 10^{-5}$  & $(75.87 \pm 3.68) \times 10^{-5}$  \\ \hline
		\end{tabular}
	}
\end{table*}

\begin{table*}
	\centering
	\caption{AIC, BIC, PSNR and RMSE for discrete wavelets transforms (WT) and undecimated wavelet transforms (UWT) with the minimum AIC.}
	\label{tab:best-wavelet}
	\resizebox{\textwidth}{!}{%
		\begin{tabular}{@{\extracolsep{3pt}}lcccccc@{}}
			\hline
			& \multicolumn{2}{c}{\textbf{Scenario 1}} & \multicolumn{2}{c}{\textbf{Scenario 2}} & \multicolumn{2}{c}{\textbf{Scenario 3}}                                                                                                            \\ \hline
			& \multicolumn{1}{c}{\textbf{WT}}         & \multicolumn{1}{c}{\textbf{UWT}}        & \multicolumn{1}{c}{\textbf{WT}}         & \multicolumn{1}{c}{\textbf{UWT}}  & \multicolumn{1}{c}{\textbf{WT}}  & \multicolumn{1}{c}{\textbf{UWT}}  \\
			\cline{2-3} \cline{4-5} \cline{6-7}
			\textbf{Best wavelet family} & \texttt{dmey}                           & \texttt{haar}                           & \texttt{sym7}                           & \texttt{haar}                     & \texttt{coif17}                  & \texttt{haar}                     \\
			\textbf{AIC}                 & $-19739.53 \pm 180.53$                  & $-18823.76 \pm 260.78$                  & $-20257.25 \pm 49.57$                   & $-20054.28 \pm 133.15$            & $-19744.98 \pm 156.91$           & $-18690.92 \pm 252.91$            \\
			\textbf{BIC}                 & $-19427.18 \pm 216.16$                  & $-16156.66 \pm 893.47$                  & $-20188.55 \pm 55.79$                   & $-19398.75 \pm 446.16$            & $-19450.36 \pm 185.29$           & $-15865.54 \pm 861.84$            \\
			\textbf{PSNR}                & $92.86 \pm 11.78$                       & $97.11 \pm 8.30$                        & $9.86 \pm 1.01$                         & $9.70 \pm 0.75$                   & $93.95 \pm 8.68$                 & $96.22 \pm 7.80$                  \\
			\textbf{RMSE}                & $(83.30 \pm 5.06)\times 10^{-5}$        & $(83.74 \pm 3.46)\times 10^{-5}$        & $(71.45 \pm 1.28) \times 10^{-5}$       & $(70.92 \pm 1.38) \times 10^{-5}$ & $(83.30 \pm 4.48)\times 10^{-5}$ & $(85.97 \pm 4.26) \times 10^{-5}$ \\ \hline
		\end{tabular}
	}
\end{table*}

\begin{table*}
	\centering
	\caption{AIC, BIC, PSNR and RMSE for delta basis function combined with discrete wavelets transforms (D+WT) and undecimated wavelet transforms (D+UWT) with the minimum AIC.}
	\label{tab:best-deltawavelet}
	\resizebox{\textwidth}{!}{%
		\begin{tabular}{@{\extracolsep{3pt}}lcccccc@{}}
			\hline
			& \multicolumn{2}{c}{\textbf{Scenario 1}} & \multicolumn{2}{c}{\textbf{Scenario 2}} & \multicolumn{2}{c}{\textbf{Scenario 3}}                                                                                                               \\ \hline
			& \multicolumn{1}{c}{\textbf{D+WT}}       & \multicolumn{1}{c}{\textbf{D+UWT}}      & \multicolumn{1}{c}{\textbf{D+WT}}       & \multicolumn{1}{c}{\textbf{D+UWT}} & \multicolumn{1}{c}{\textbf{D+WT}} & \multicolumn{1}{c}{\textbf{D+UWT}} \\
			\cline{2-3} \cline{4-5} \cline{6-7}
			\textbf{Best wavelet family} & \texttt{db7}                            & \texttt{db34}                           & \texttt{dmey}                           & \texttt{haar}                      & \texttt{dmey}                     & \texttt{db38}                      \\
			\textbf{AIC}                 & $-20231.67 \pm 61.31$                   & $-19931.31 \pm 204.32$                  & $-20246.30 \pm 57.76$                   & $-19984.59 \pm 168.32$             & $-20104.92 \pm 70.98$             & $-19365.93 \pm 274.70$             \\
			\textbf{BIC}                 & $-20149.23 \pm 68.95$                   & $-19110.59 \pm 660.77$                  & $-20134.18 \pm 70.59$                   & $-17275.06 \pm 914.83$             & $-19937.21 \pm 82.66$             & $-16832.39 \pm 1064.69$            \\
			\textbf{PSNR}                & $123.42 \pm 6.28$                       & $123.11 \pm 6.22$                       & $10.00 \pm 0.95$                        & $9.64 \pm 0.72$                    & $116.75 \pm 7.20$                 & $117.38 \pm 6.58$                  \\
			\textbf{RMSE}                & $(71.98 \pm 1.55)\times 10^{-5}$        & $(72.47 \pm 1.96)\times 10^{-5}$        & $(71.31 \pm 1.49) \times 10^{-5}$       & $(71.47 \pm 1.32) \times 10^{-5}$  & $(74.45 \pm 1.85)\times 10^{-5}$  & $(74.60 \pm 2.13) \times 10^{-5}$  \\ \hline
		\end{tabular}
	}
\end{table*}

\begin{figure*}
	\centerline{\includegraphics[width=\linewidth]{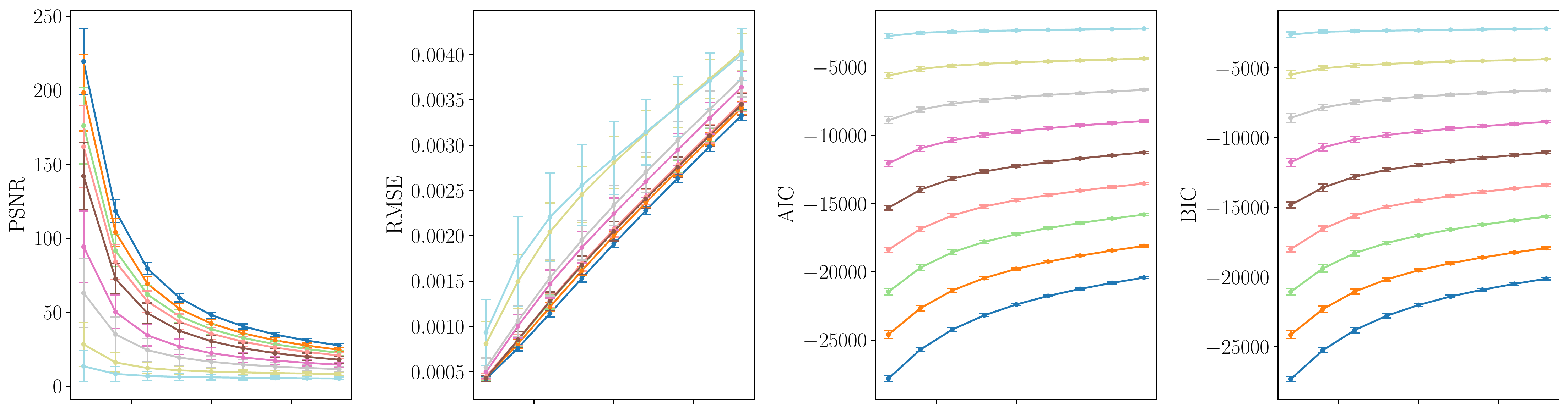}}
	\centerline{\includegraphics[width=\linewidth]{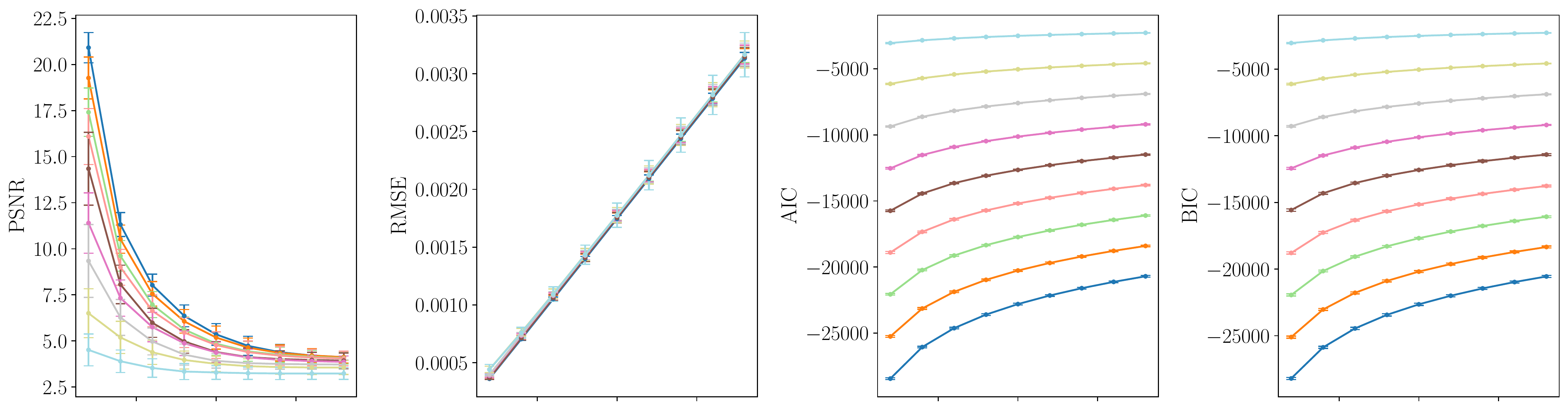}}
	\centerline{\includegraphics[width=\linewidth]{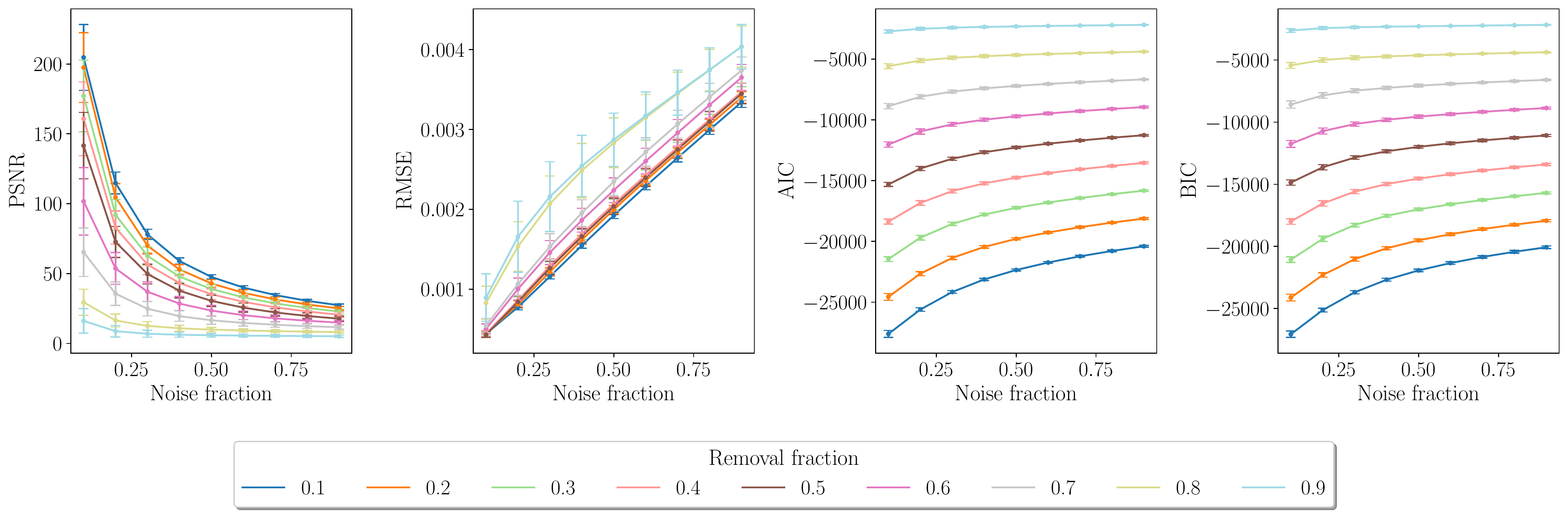}}
	\caption{PSNR, RMSE, AIC and BIC using a discrete wavelet basis reconstruction. The first row shows Scenario~1, the second row shows Scenario~2 and the third row shows Scenario~3. We have used wavelets \texttt{dmey}, \texttt{sym7} and \texttt{coif17}, respectively.}
	\label{fig:csbasis_aicbic}
\end{figure*}

Given the fact that the deconvolution of Faraday depth spectra is an ill-posed problem, many models will be able to fit the data. Additionally, since we are using multiple scenarios to test different basis representations, we would like to know which model best approximates the data. To do this we use the Akaike Information Criterion (AIC) and the Bayesian Information Criterion (BIC) defined as
\begin{equation}
	\text{AIC} = n \log{\biggl(\frac{||y - \hat{y}(x)||^2_2}{n}\biggr)} + 2\cdot \text{df},
\end{equation}
and
\begin{equation}
	\text{BIC} = n \log{\biggl(\frac{||y - \hat{y}(x)||^2_2}{n}\biggr)} + \text{df} \log{n},
\end{equation}
respectively. Here, $y$ represents the complex observed data vector of length $n$ and $\hat{y}$ represents the estimated data, and $x$ can be either the Faraday depth spectrum or the coefficient vector of length $n$. Following \cite{SHI201872}, $\text{df}$ is an estimate of the degrees of freedom of $x$, which is defined as the number of non-zero entries.

Although in \cite{SHI201872} AIC and BIC have been used as a part of the optimisation process, here we calculate them using the estimated data only after convergence. In summary, the only difference between the AIC and BIC is that the BIC takes into account the logarithm of the total size of the estimator. Here we are looking for the sparser signal that best minimizes the residual sum of squares (RSS). In that sense, in the BIC the function will also be penalized if the size of the signal is large. However, since we are testing for different percentages of removal fraction, which will remove data randomly in $\lambda^2$-space, this can change the size of the Faraday depth space and consequently the size of the wavelet representation. Therefore, we suggest that $n$ should be equal to the number of measurements in $\lambda^2$-space, such that the AIC and BIC metrics show the sparsest signal with the best RSS. Further information about the use of the AIC and BIC can be found in \cite{aicbicreview}.

In order to study which wavelet family best represents certain scenarios we have made 50 realizations for each of the scenarios described in Section~\ref{sec:simdata}, fixing the noise and removal fractions to $0.7$\;mJy/beam and $0.3$, respectively. This allows us to calculate the mean and standard deviation of each metric. We have used only orthogonal wavelet families and  the delta function basis. It is important to highlight that when using wavelet transforms we always use the maximum number of decomposition levels defined for the DWT. For a signal of length $n$ and a filter of length $p$ this is defined as
\begin{equation}
	\text{max\_level} = \biggl\lfloor \log_2{\biggl(\frac{n}{p-1}\biggr)} \biggr \rfloor \;.
\end{equation}
The results of this process for the DWT is shown in Figure~\ref{fig:wavelet_study}. For the corresponding UWT the number of decomposition levels is given by $\log_2(n)$ and the results of these experiments are shown in Figure~\ref{fig:undecimated_wavelet_study} for undecimated wavelets.

From Figure~\ref{fig:wavelet_study} and Figure~\ref{fig:undecimated_wavelet_study} it can be seen that although the average values of the RMSE, AIC and BIC tend to be lowest for the delta function basis, the best values for wavelets can be obtained using the discrete Meyer, Haar, symlets and coiflets filters. However, when combining the wavelets with the delta function basis the best metrics are obtained using the discrete Haar and Daubechies, Meyer and Haar families.

Note that although AIC and BIC tend to be higher for UWT, one must take into account that the number of coefficients are larger than in the discrete wavelet transform case since the UWT does not use any downsampling or upsampling.

From these results we have selected the best wavelet families to be those that have the lowest AIC and BIC average values. These are summarised in Tables~\ref{tab:best-family} and ~\ref{tab:best-deltafamily} where all the metrics for the three scenarios and wavelet and undecimated transforms are presented. Then we have repeated the same process by selecting the best wavelets within those families for each of the cases. These results are presented in Tables~\ref{tab:best-wavelet} and \ref{tab:best-deltawavelet}. As a comparison we also present these results for the delta function basis in Table ~\ref{tab:deltafunctionbasis}. Note that even though the WT works better than UWT according to the AIC and BIC, the difference is small. The tests show that PSNR and RMSE are improved when using UWT. This can be attributed to the fact that the UWT is considered a multiscale decomposition \citep{starck}.

\subsubsection{Data corruption evaluation}
\label{sec:corruption}

Having selected our preferred wavelet bases, as described in Section~\ref{sec:aic}, we make 50 different realizations of the reconstructions for each of the three scenarios using a delta function basis, a discrete wavelet basis (DWT) and an undecimated wavelet basis (UWT), respectively. The first thing that we note for all cases and scenarios is that as the noise and removal fractions increase the PSNR decreases; see Figure~\ref{fig:csbasis_aicbic}. This means that the algorithm struggles to recover the signal as it starts increasingly to fit the noise. The RMSE increases as well, which means that the residuals start to vary more as a function of $\lambda^2$. This is translated into a noisier model in Faraday depth.

From the AIC and BIC results for the delta function and discrete wavelet cases we can see that as noise increases the sparsity of the signals decrease, see Figure~\ref{fig:csbasis_aicbic} (and Figure~\ref{fig:deltafbasis_aicbic}). For the delta function basis, this means that the model is fitting more noise; however, for the wavelet basis this means that more coefficients are needed in order to fit a noisier signal. Additionally, we note that errors increase as noise and removal fraction increase, illustrating, as expected, that the reconstructions are sensitive to these two factors.

Equivalent results for the UWT are shown in Figure~\ref{fig:undecimated_csbasis_aicbic}, where very different outcomes compared to the discrete wavelet transform, see Figure~\ref{fig:csbasis_aicbic}, can be seen. Even though the PSNR and RMSE plots look similar, the AIC and BIC plots do not. Care must be taken when using these latter metrics and the UWT. Not only do the number of elements increase (due to not decimating the coefficients), but as do the number of non-zero elements (degrees of freedom). These tend to be $\sim$~100 times more than in the discrete wavelet and therefore they often surpass the number of parameters in $\lambda^2$-space.

\section{Application to Real Data}
\label{sec:realdata}

\subsection{Abell 1314}

Abell\,1314 (A1314; $z=0.034$) is a nearby low-mass merging galaxy cluster with a highly disturbed density profile. The observational properties of this cluster are summarised in Table~\ref{tab:a1314}. From shallow \emph{XMM-Newton} observations, \cite{wilber2019} calculate the $0.5-2.4$\,keV X-ray luminosity of A1314 to be $0.17 \times 10^{44}$\,ergs\,s$^{-1}$ (higher than that previously found by \cite{ledlow2003} using ROSAT data) and hence derived a mass of $M_{500} = 0.68 \times 10^{14}$\,M$_{\odot}$ using the scaling relation of \cite{reichert2011}.

Although no deep observations of A1314 have been made in the X-ray, a significant number of radio studies have looked at the cluster due to the complex nature of the radio emission associated with the cluster galaxies. The brightest of these, IC\,708 and IC\,711, are bent tail galaxies, and IC\,711 in particular has a very long radio tail extending for almost a mega-parsec in a SE-NW direction across the cluster \citep{vallee1976}. Observations from 2016 with the GMRT \citep{srivastava2020} showed a spectral break part of the way along the tail of IC\,711, and suggested that this may arise due to local environmental factors within the cluster. This was supported by later observations, which showed low-levels of diffuse emission  perpendicular to the main direction of the IC\,711 tail, again indicating a disturbance within A1314 \citep{sebastian2017}.

Despite the disturbed nature of the cluster, recent radio observations with the LOFAR and GMRT telescopes did not detect a radio halo \citep{wilber2019}, but the authors suggest that this may be because of the very low mass of this system. However, they do detect irregularly-shaped diffuse radio emission likely to be associated with historic AGN activity from the IC\,712. They also note that the spectral index map of the extended head-tail radio galaxy IC\,711 indicates signs of disturbance, consistent with the findings of \cite{srivastava2020}.

\begin{table}
	\caption{Observational properties of Abell\,1314 from the literature. \label{tab:a1314}}
	\resizebox{0.48\textwidth}{!}{%
		\begin{tabular}{lll}
			\hline
			Property         & Value                                 & Reference          \\\hline
			R.A. (J2000)     & 11 34 48.7                            & \cite{mahdavi2001} \\
			$\delta$ (J2000) & +49 02 25                             & \cite{mahdavi2001} \\
			$z$              & 0.034                                 & \cite{popesso2004} \\
			$L_{\rm X}$      & $0.17 \times 10^{44}$\,ergs\,s$^{-1}$ & \cite{wilber2019}  \\
			$M_{500}$        & $0.68 \times 10^{14}$\,M$_{\odot}$    & \cite{wilber2019}  \\\hline
		\end{tabular}
	}
\end{table}

Here we use a combination of archival X-ray and radio observations to derive the magnetic field properties of A1314 as a function of location within the cluster.

\subsection{X-ray Data: Calibration and Imaging}

A1314 was observed by \emph{XMM-Newton} in November 2003 during rev. 725 with a total exposure time of 18.4\,ks. Since the primary focus of this paper is the analysis of radio data we do not provide a summary of the X-ray processing performed in this work here; however, a full description can be found in Appendix~\ref{app:xrayreduction}. The resulting X-ray surface brightness distribution is shown in Figure~\ref{fig:a1314_overlay}.

We derive the X-ray temperature of A1314 from spectral fitting of the \emph{XMM-Newton} data to be $T_{\rm X} = 2.2$\,keV on average for the cluster, broadly consistent with that found by \cite{wilber2019}. A radial temperature profile showing a cooler core region can be found in Figure~\ref{fig:temp}.

\subsection{Radio Data: Calibration and Imaging}

In this work we use radio data from the VLA telescope towards A1314. The three hour observation was made on January 2019 in L-Band (1.008-2.031\,GHz) and using the C~configuration. This information is summarised in Table~\ref{tab:JVLA-obs}.
%
\begin{table}
	\caption{Abell\,1314 VLA radio observation details.}
	\label{tab:JVLA-obs}
	\resizebox{0.48\textwidth}{!}{%
		\begin{tabular}{@{}lccccccr@{}}
			\hline
			Obs. Date    & Obs. Time & Band & VLA   & $\nu_{\text{ref}}$ & Beam                 & $\sigma_{\rm rms}$ \\
			& [hr]      &      & Array & [GHz]              & [arcmin]             & [mJy/bm]           \\\hline
			January 2019 & 3         & L    & C     & 1.5                & $0.21' \times 0.20'$ & 0.09               \\\hline
		\end{tabular}
	}
\end{table}

\begin{table}

	\caption{A1314 radio sources. Radio sources within the A1314 field detected in Stokes~I at $\ge 5\,\sigma_{\rm I}$ are listed in order of increasing Right Ascension. Column [1] lists the source id as used in this work, [2] Right Ascension of the source in degrees, [3] Declination of the source in degrees, [4] peak flux density of the source at a frequency of 1.5\,GHz, [5] Stokes~I spectral index of the source at the position of the peak as determined from the VLA data used in this work, [6] distance of the source from the X-ray centre in kpc, and [7] indicates whether the source is also detected in polarisation at $\ge 6\,\sigma_{\rm P}$.}
	\label{tab:a1314-sources}
	\begin{threeparttable}[b]
		\begin{tabular}{cccrcrc}
			\hline
			ID           & \multicolumn{1}{c}{R.A}   & \multicolumn{1}{c}{Dec}   & \multicolumn{1}{c}{$S_{\rm peak}$} & \multicolumn{1}{c}{$\alpha$} & \multicolumn{1}{c}{Distance} & \multicolumn{1}{c}{Pol?} \\
			& \multicolumn{1}{c}{(deg)} & \multicolumn{1}{c}{(deg)} & \multicolumn{1}{c}{(mJy/bm)}       &                              & \multicolumn{1}{c}{(kpc)}    & [Y/N]                    \\\hline
			1            & 173.454                   & 48.987                    & 9.11                               & -0.73                        & 520.2                        & Y                        \\
			2            & 173.496                   & 49.046                    & 32.04                              & -0.72                        & 390.7                        & N                        \\
			3\tnote{a}   & 173.497                   & 49.062                    & 156.69                             & -0.43                        & 378.9                        & Y                        \\
			4            & 173.498                   & 48.941                    & 2.57                               & 0.12                         & 538.7                        & N                        \\
			5            & 173.521                   & 49.106                    & 4.24                               & -0.82                        & 331.5                        & Y                        \\
			6            & 173.571                   & 49.152                    & 14.11                              & -0.53                        & 291.1                        & N                        \\
			7            & 173.621                   & 48.951                    & 1.29                               & -1.83                        & 397.6                        & N                        \\
			8            & 173.632                   & 49.048                    & 1.59                               & 0.57                         & 179.4                        & N                        \\
			9 \tnote{b}  & 173.694                   & 48.956                    & 41.27                              & -0.22                        & 353.0                        & Y                        \\
			10 \tnote{c} & 173.705                   & 49.078                    & 27.97                              & -0.72                        & 36.9                         & Y                        \\
			11           & 173.713                   & 49.203                    & 1.87                               & -1.20                        & 291.0                        & N                        \\
			12           & 173.761                   & 49.193                    & 4.29                               & -0.63                        & 277.6                        & Y                        \\
			13           & 173.803                   & 48.967                    & 11.17                              & -0.98                        & 357.6                        & Y                        \\
			14           & 173.933                   & 49.038                    & 3.62                               & -0.68                        & 400.0                        & N                        \\
			15           & 173.942                   & 48.921                    & 3.55                               & 0.41                         & 590.1                        & Y                        \\ \hline
		\end{tabular}
	\end{threeparttable}
	\begin{tablenotes}
		\item [a] IC708.
		\item [b] IC711.
		\item [c] IC712.
	\end{tablenotes}
\end{table}

\begin{figure*}
	\centering
	\includegraphics[width=\textwidth]{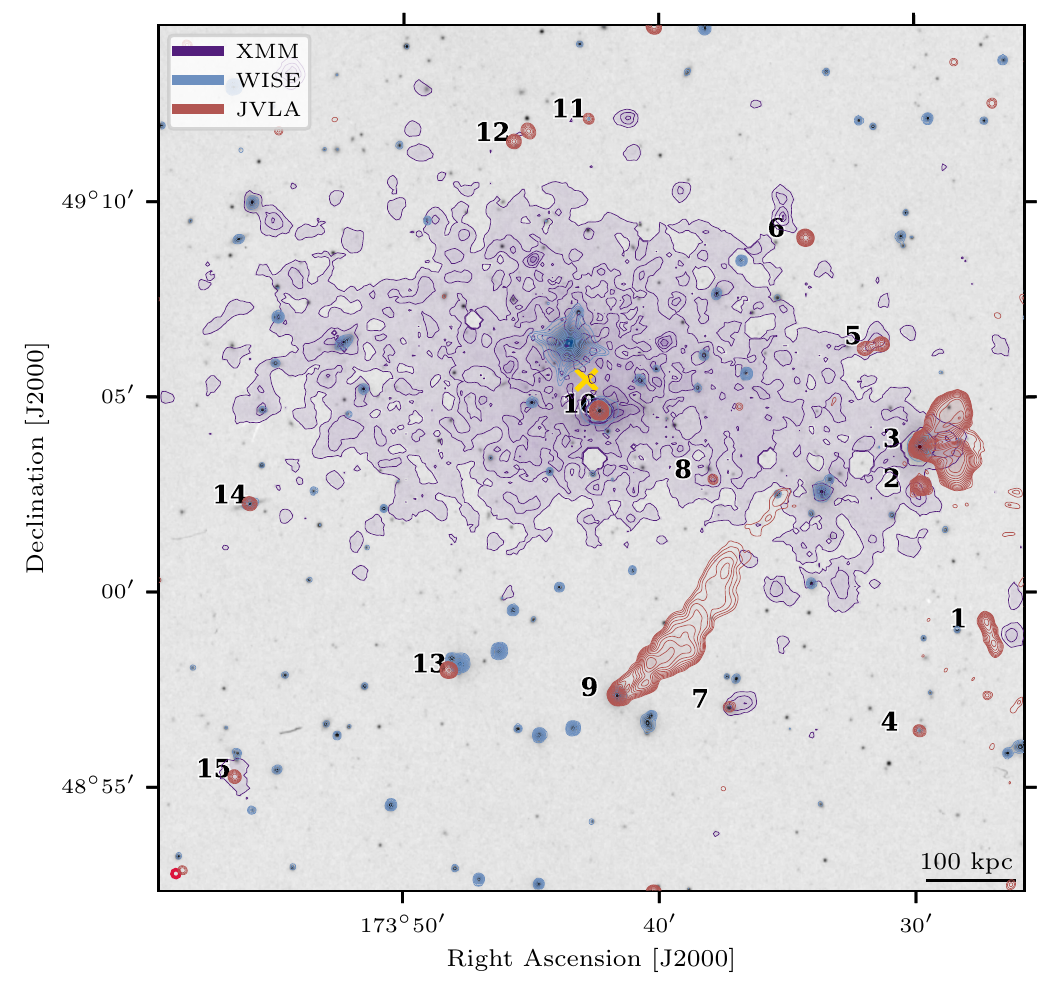}
	\caption{Overlay image of cluster Abell 1314. Image of the SDSS optical spectral with three overlayed contours. Purple contours represent the X-ray XMM-Newton point-source subtracted surface brightness. Blue contours represent the WISE 3.4\,$\mu$m infrared emission. Finally, red contours represent the VLA C-configuration total intensity radio emission at 1.5 GHz. The contours start at $3.75\sigma$, $50\sigma$ and $5.0\sigma$, respectively and all of them are spaced by a factor of 2. $\sigma$ value for radio contours is listed in Table \ref{tab:JVLA-obs}. The eleven detected radio galaxies are marked with numbers and the cluster center is marked with a yellow cross. See Table~\ref{tab:a1314-sources} for further information about each source.}
	\label{fig:a1314_overlay}
\end{figure*}

The data were manually inspected before parallel-hand calibration in order to find anomalies and/or RFI. Subsequently the VLA pipeline was used to calibrate the parallel-hand data. To use the pipeline we followed the polarization calibration {\sc casa} guide\footnote{\url{https://casaguides.nrao.edu/index.php?title=CASA_Guides:Polarization_Calibration_based_on\_CASA_pipeline_standard_reduction:_The_radio_galaxy_3C75-CASA5.6.2}}. Following the pipeline calibration, the Measurement Set was then restored to its state before polarization calibration in order to use our custom polarization calibration framework.

Before doing polarization calibration we inspected the amplitude-frequency plots of the cross-hand data for each spectral window in order to identify RFI and instrumental anomalies in the data. These were flagged using the automated statistical algorithms {\tt rflag} and {\tt tfcrop} in {\sc casa}. The first of these is used to remove the residual RFI and although the use of {\tt tfcrop} is optional, we use it here to decrease the amount of residual RFI in the parallel-hand data. We note that the RFI removal works better over short frequency intervals (no more than 3 spectral windows). We first flag the residual RFI in the parallel hands followed by the cross-hand data.

By considering the available calibrators in the observation, we chose 3C286 as the polarization angle calibrator
and the unpolarized source 3C147 as the leakage calibrator. We then applied a four-step polarization calibration framework: fitting the calibrator data to known flux density, polarization angle and polarization fraction models, calibrating for cross-hand delays, calibrating for leakage, and lastly calibrating for polarization angle.

The polarization calibration framework begins by using the Taylor coefficients of 3C286 and 3C147 and their errors as stored in {\sc casa} from \cite{Perley_2017} to calculate  flux densities for each source as a function of frequency. Using the {\sc casa} task {\tt setjy} we then create Stokes IQUV point source models (Stokes V is set to zero) for 3C286 and 3C147. Note that 3C147 is unpolarized at these frequencies and therefore Stokes Q, U and V will be equal to zero for this calibrator.

The framework continues with the cross-hand delay calibration, which solves cross-hand delays due to the residual difference between correlations R and L on the reference antenna. We use 3C286 as the cross-hand delay calibrator since we know that it has polarized signal in the RL and LR correlations. There are two ways to solve for cross-hand delays. The first option is to solve for a multiband delay, which fits the cross-hand delay across the entire baseband. The second option is to solve the cross-hand delays independently for each spectral window. In this case, we choose the first option. This is performed using the {\sc casa} task \texttt{gaincal} with the parameter {\tt gaintype='KCROSS'} to create the cross-hand delay calibration solution table.

Using the unpolarized calibrator 3C147 it is possible to derive the polarization leakage D-terms for the VLA dataset. This is done using the {\sc casa} task {\tt polcal} to derive a frequency-dependent leakage correction assuming zero intrinsic polarization. We follow this with a frequency-dependent polarization angle calibration on 3C286 using the {\sc casa} task {\tt polcal}.

Following calibration for parallel-hands and cross-hands, we image the target using multi-scale multi-frequency synthesis (MFS) using the {\sc casa} task {\tt tclean} with 128 w-projection planes. We make Stokes I, Q and U images of the entire field, $22 \times 22$\,arcminutes, with a cell size of $1.3$\,arcseconds. Finally, we use {\tt widebandpbcor} to correct for the primary beam. The calibrated total intensity image has a noise level of 0.09\,mJy/beam and a peak flux density of 156.69\,mJy/beam. Self-calibration was not applied in this instance.

Figure~\ref{fig:a1314_overlay} shows the SDSS optical image of cluster Abell 1314 overlaid with the XMM-Newton data (purple), WISE 3.4\,$\mu$m (blue) and VLA radio (red) contours. Eleven radio sources are detected above a level of 5\,$\sigma_{\rm rms}$ in Stokes~I. Two radio galaxies (marked with IDs 3 and 9) are significantly resolved in this configuration. The brightest radio galaxy in the field, also known in the literature as IC708, has a peak flux density of 156.69\,mJy/beam, shows a wide angle tail, and is located at a projected distance of 378.9\,kpc from the X-ray centre (marked with a yellow cross). The second, known also as IC711, has a peak flux of 41.27\,mJy/beam and in this configuration it is possible to see part of the very long radio tail extending to the north-west of the source. The host of this radio galaxy is located at a projected distance of 353.0\,kpc from the X-ray centre. Information on all radio sources can be found in Table~\ref{tab:a1314-sources}. Of these sources we find that eight sources have detectable polarized emission.

To produce Stokes I, Q and U cubes in preparation for RM Synthesis, we use the {\sc casa} multi-scale {\tt tclean} task with option {\tt specmode='cubedata'} to create multi-frequency spectral cubes. As described in Section~\ref{sec:spectral}, we restore all frequency channels using a PSF equal in size to that of the MFS image, see Table~\ref{tab:JVLA-obs}.

\begin{table}
	\begin{center}
		\caption{Abell\,1314 VLA resolution and field-of-view details \label{tab:jvla-params}}
		\begin{tabular}{@{}crl@{}}
			\hline
			Parameter              & Value                 & Description        \\\hline
			$\lambda_{\text{min}}$ & $0.148\;\text{m}$     & Minimum wavelength \\
			$\lambda_{\text{max}}$ & $0.297\;\text{m}$     & Maximum wavelength \\
			$B_{\text{max}}$       & $3373.853\;\text{m}$  & Maximum baseline   \\
			$D_{\text{min}}$       & $25\;\text{m}$        & Dish size          \\
			$\Delta x$             & $9.02\;\text{arcsec}$ & Resolution         \\
			FOV                    & $40.8\;\text{arcmin}$ & Field-of-view      \\\hline
		\end{tabular}
	\end{center}
\end{table}

\subsection{Faraday depth Reconstruction}
\label{sec:reconstruction}

\begin{table*}
	\centering
	\caption{Different noise/variance quantities used in data reconstruction and analysis.  \label{tab:noise}}
	\begin{tabular}{l|l|l}
		\hline
		Noise quantity          & Definition                               & Method of measurement/calculation                                                           \\\hline
		$\sigma_{\rm I}$        & Noise in integrated Stokes~I             & Measured from MFS Stokes~I image using the rms in an off-source region                      \\
		$\sigma_{\rm P}$        & Noise in integrated Stokes~P             & Measured from MFS Stokes~Q~\&~U images using the rms in an off-source region.               \\
		$\sigma_{\rm Q}$        & Channel noise in Stokes~Q                & Measured from each Stokes~Q channel image using the rms in an off-source                    \\
		&                                          & region                                                                                      \\
		$\sigma_{\rm U}$        & Channel noise in Stokes~U                & Measured from each Stokes~U channel image using the rms in an off-source                    \\
		&                                          & region                                                                                      \\
		$\sigma_{\rm QU}$       & Arithmetic mean of QU channel noise      & Calculated from $\sigma_{\rm Q}$ and $\sigma_{\rm U}$, see Section~\ref{sec:reconstruction} \\
		$\sigma_{\phi}$         & Theoretical noise in Faraday depth space & Reciprocal of the square root of the sum of the frequency channel weights,                  \\
		& (real/imaginary parts)                   & see Section~\ref{sec:reconstruction}.                                                       \\
		$\sigma'_{\phi}$        & Measured noise in Faraday depth space    & Measured as the rms of the reconstructed Faraday depth spectrum in the                      \\
		& (real/imaginary parts)                   & regions $|\phi| > 0.8\,\phi_{\rm max}$, see Section~\ref{sec:reconstruction}.               \\
		$\sigma_{\rm RM}$       & Faraday dispersion                       & See Equation~\ref{eq:sigma_rm}                                                              \\
		$\Delta\phi_{\rm peak}$ & Faraday uncertainty at peak              & See Equation~\ref{eq:delta_phi_peak}                                                        \\\hline
	\end{tabular}
\end{table*}

Before RM-Synthesis we need to prepare the data and look for outliers that might affect the quality of the reconstruction. Firstly, all the channels from the resulting cubes are checked by {\tt cs-romer}, following which 33.79\% of the frequency slices from the cubes were discarded since they resulted in empty images. This is mainly due to frequency channels flagged during parallel and cross hand calibration. We calculate the rms noise in each slice of the Q and U cubes, resulting in $\sigma_Q$ and $\sigma_U$ values for each channel, see Table~\ref{tab:noise}. Assuming $\sigma_Q \approx \sigma_U$, we define $\sigma_{QU} = \frac{\sigma_Q + \sigma_U}{2}$. From this, we calculate the mean and the standard error of $\sigma_{\text{QU}}$ as $\langle{\sigma_{QU}}\rangle$ and $\hat{\sigma}_{{QU}}$, respectively. Subsequently, we flag an additional 12.01\% of the data slices that have $\sigma_{\rm QU} > \langle{\sigma_{QU}}\rangle + 5\hat{\sigma}_{QU}$. Finally, we select those lines of sight where both total intensity and polarized intensity in the MFS images are greater than 3$\sigma_{\rm I}$ and 3$\sigma_{\rm P}$, respectively. The quantity $\sigma_{\rm P}$ is the noise in integrated Stokes~P, which has a Ricean distribution with a standard deviation equivalent to the rms in integrated Stokes~Q and U. We denote this quantity $\sigma_{\rm P}$ to differentiate it from the channel noise, $\sigma_{\rm QU}$, see Table~\ref{tab:noise}. This results in a total of 678 channels and $16,280$ unflagged lines of sight to be reconstructed.

\begin{table}
	\begin{center}
		\caption{Abell\,1314 VLA RM-Synthesis details. \label{tab:rmsynthesis-params}}
		\begin{tabular}{@{}lcccc@{}}
			\hline
			\textbf{Parameter}             & $\delta \phi$ & max-scale & $|\phi_{\text{max}}|$ & $\phi_R$ \\\hline
			\textbf{Value [rad\,m$^{-2}$]} & 54.4          & 129.7     & 13540.6               & 6.8      \\\hline
		\end{tabular}
	\end{center}
\end{table}

We run {\tt cs-romer} over the $P(\lambda^2)$ cube using weights calculated directly from each slice as $W(\lambda^2) = 1/{\sigma^2_{QU}(\lambda^2)}$. The framework then calculates a Faraday depth reconstruction with parameters as shown in Table~\ref{tab:rmsynthesis-params}.

The stopping (convergence) criterion for {\tt cs-romer} is determined by the $\eta$ parameter defined in Equation~\ref{eq:eta}. In the case of real data, where $\sigma_{\phi}$ is unknown a priori, there are multiple possible approaches to calculate this quantity. The first is to use the measured noise from the Stokes~Q and U frequency channels and to calculate
\begin{equation}
	\sigma_{\phi} = \Biggl[\sqrt{\sum_{i=1}^{N} W_i(\lambda^2)}\Biggr]^{-1},
\end{equation}
where $N$ is the number of unflagged frequency channels; alternatively one might estimate $\sigma_{\phi}$ directly from the dirty Faraday depth spectrum towards each polarisation detection. This second approach is best done using the edges of the spectrum, which are expected to contain least contamination due to sidelobe structure; however, tests using simulated data as described in Section~\ref{sec:simdata} suggest that $\sigma_{\phi}$ calculated directly from the Faraday depth spectrum will always be over-estimated when significant structure is present, regardless of whether it is measured using a direct rms calculation or using the median absolute deviation, which is in principle more robust to outliers such as those from residual sidelobe structure. Therefore for the reconstruction in this work we calculate $\sigma_{\phi}$ using the first of these methods and we note that this approach gives $\sigma_{\phi}$ values consistent with the empirically measured rms recovered from empty, i.e. structure-free, lines of sight in Faraday depth.

\begin{figure}
	\centering
	\includegraphics[width=0.5\textwidth]{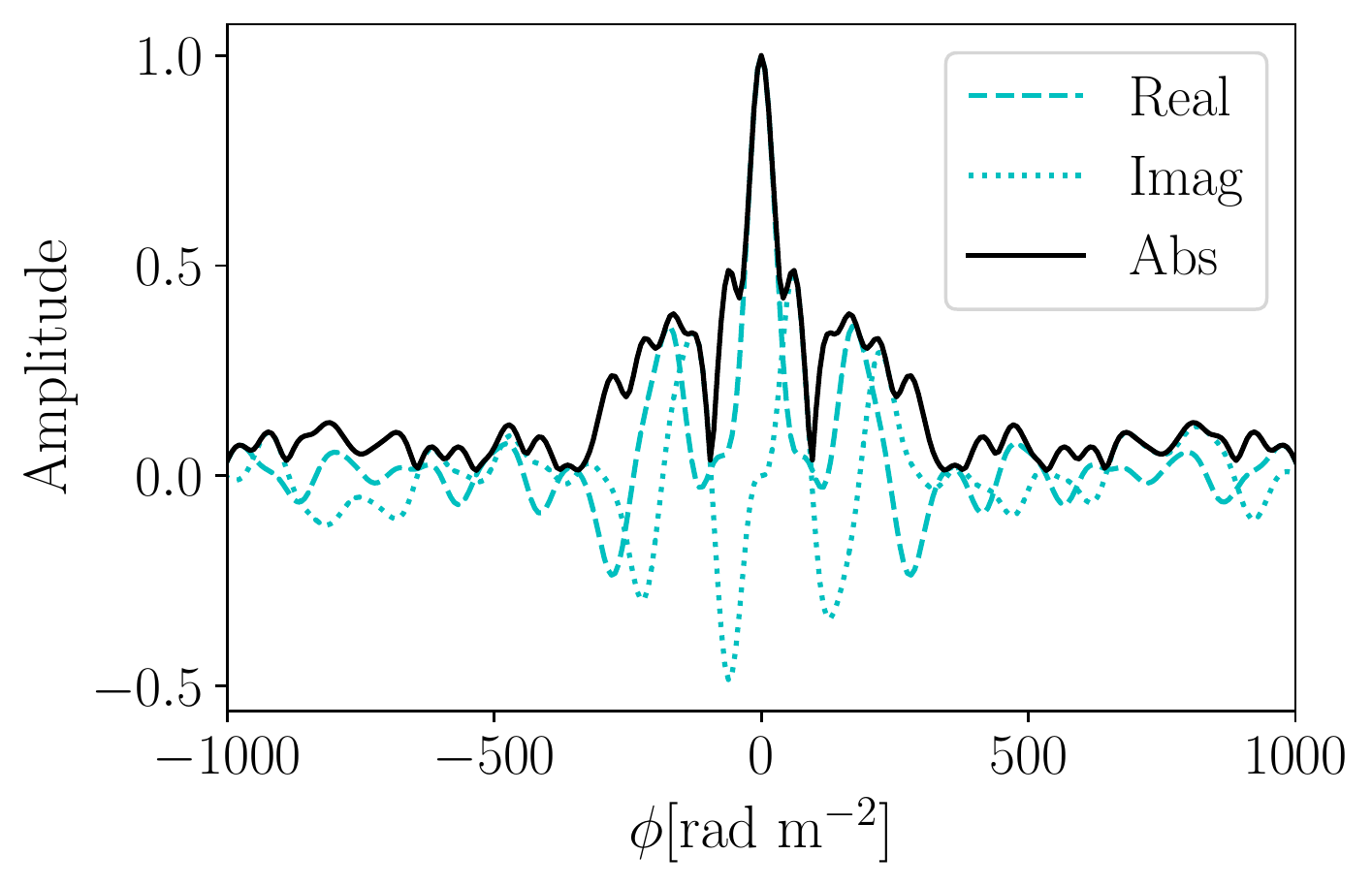}
	\caption{RMTF for the VLA Abell 1314 data.}
	\label{fig:rmtf}
\end{figure}

Unlike \cite{stuardi2021}, who use an average spectral index of $\alpha=1$ for all lines of sight, we use the {\tt tclean} MFS spectral index image to reconstruct Faraday depths. Following the results described in Section~\ref{sec:aic} and shown in Table~\ref{tab:deltafunctionbasis}, we use the delta basis function for reconstruction. Faraday depth spectra are recovered for $16,280$ lines of sight over the range $[-\phi_{\text{max}},\phi_{\text{max}}]$. The RMTF for these data is shown in Figure~\ref{fig:rmtf}.

We compare the rms of the residuals for each reconstructed line of sight to the theoretical value of $\sigma_{\phi}$ calculated from the weights on the spectral data. Unlike the reconstructions of the simulated scenarios in Section~\ref{sec:simdata}, for some lines of sight the measured rms differs from the theoretical value, and we suggest that this is likely due to residual RFI contamination in the real VLA data. We find that the rms of the residuals for the VLA data is consistent with that calculated using the edges of the clean Faraday depth spectrum, which we estimate using the regions where $|\phi| > 0.8\,\phi_{\rm max}$ and refer to as $\sigma'_{\phi}$, see Table~\ref{tab:noise}.

The {\tt cs-romer} framework outputs the following data products: a dirty Faraday depth cube, a Faraday depth model cube, a Faraday depth residual cube and a restored Faraday depth cube. RM images for each polarised source are shown in Figures~\ref{fig:ic708}\,-\,\ref{fig:ic712}. These show the peak Faraday depth, $\phi_{\text{peak}}$, the uncertainty at the peak, $\Delta\phi_{\rm peak}$, calculated as
\begin{equation}
	\Delta\phi_{\rm peak} = \frac{\delta \phi}{2P/\sigma'_{\phi}}\;,
	\label{eq:delta_phi_peak}
\end{equation}
and the fractional polarization, using polarized intensity corrected for Ricean bias such that
\begin{equation}
	P = \sqrt{|F(\phi_{\text{peak}})|^2 - 2.3{\sigma'}_{\phi}^{2}}
\end{equation}
\citep{george_stil_keller_2012}, where $\sigma'_{\phi}$ is calculated as described above. The polarization fraction is calculated by dividing $P$ by the Stokes I MFS image in regions where the emission exceeds both a $3\sigma_{P}$ and $3\sigma_I$ threshold.

\subsection{Galactic RM contribution}
\label{sec:grm}

The mean RM of the Galactic contribution for the Abell 1314 field is $\langle \phi_{\rm GAL} \rangle = -15 \pm 6$\;rad/m$^2$ \citep{faradaysky2020}. Since this is systematically different from zero, we choose to subtract it from the VLA measurements. This is performed using the {\tt cs-romer} framework, as described in Section~\ref{sec:derotgal}, by querying the Galactic foreground RM mean and standard deviation maps provided by \cite{faradaysky2020}. These foreground maps are significantly lower resolution than the JVLA data, and cover the A1314 field using $\sim$13 pixels. The {\tt cs-romer} framework uses a bilinear interpolation across the field-of-view to approximate the Galactic foreground at the position of each pixel in the VLA data. The mean Galactic RM values for each line-of-sight are then derotated as a shift to each line of sight for the polarized intensity in $\lambda^2$ space as described in Section~\ref{sec:derotgal} using Equation \ref{eq:gal-derot}.

\subsection{Polarised radio galaxies}
\label{sec:radgal}

Eight radio sources in the A1314 field have detectable polarised emission above a level of $6\sigma_{P}$, see Table~\ref{tab:a1314-sources}. For each of these sources we calculate the pixel-wise average rotation measure across the source, $\langle \text{RM} \rangle$, the observed standard deviation of the RM across the source, $\sigma_{\rm RM, obs}$, and the median error of the RM estimate, ${\rm med}(\sigma'_{\phi})$. The true dispersion of the RM across the source is then calculated as
\begin{equation}
	\label{eq:sigma_rm}
	\sigma_{\rm RM} = \sqrt{\sigma_{\rm RM, obs}^2 - {\rm med}({\sigma'}_{\phi})^2}.
\end{equation}

The value of $\sigma_{\rm RM}$ is shown for each source in Table~\ref{tab:a1314_pol}, along with the median RM for each source, ${\rm med}(\text{RM})$, and the median absolute deviation (MAD) value, which is considered to provide an improved estimate of the true dispersion when outliers are present in the data. Sources which cover a sky area smaller than 5 synthesized beams do not have dispersion, standard deviation or MAD values calculated due to an insufficiency of independent samples.

\begin{table*}
	\caption{RM profile of Abell 1314 polarized sources.}
	\label{tab:a1314_pol}
	\begin{threeparttable}[b]
		\centering
		\begin{tabular}{@{}ccccccccccc@{}}
			\hline
			ID & z       & $\langle \text{RM} \rangle$ & $\langle \text{RM} \rangle$\tnote{a} & $\sigma_{\text{RM}}$ & med(RM)        & MAD(RM)        & med($\sigma_{\phi}$) & $n_{\text{beam}}$ \\
			&         & (rad m$^{-2}$)              & (rad m$^{-2}$)                       & (rad m$^{-2}$)       & (rad m$^{-2}$) & (rad m$^{-2}$) & (rad m$^{-2}$)       &                   \\\hline
			1  & $-$     & 5.5                         & 19.4                                 & 57.3                 & 20.5           & 6.8            & 1.6                  & 7.8               \\
			3  & $0.032$ & -6.7                        & 10.7                                 & 58.5                 & $-$            & 27.3           & 0.9                  & 62.5              \\
			5  & $-$     & 26.8                        & 45.2                                 & $-$                  & 34.1           & $-$            & 1.5                  & 2.0               \\
			9  & $0.032$ & 11.1                        & 4.8                                  & 61.7                 & 6.8            & 13.7           & 1.3                  & 84.4              \\
			10 & $0.033$ & 19.1                        & 35.9                                 & $-$                  & 34.1           & $-$            & 1.6                  & 1.4               \\
			12 & $-$     & -68.3                       & -51.6                                & $-$                  & -54.6          & $-$            & 1.7                  & 1.1               \\
			13 & $-$     & 6.7                         & 20.4                                 & $-$                  & 20.5           & $-$            & 1.0                  & 1.8               \\
			15 & $-$     & -25.3                       & -15.8                                & $-$                  & -20.5          & $-$            & 4.1                  & 0.4               \\\hline
		\end{tabular}
		\begin{tablenotes}
			\item [a] Galactic-contribution-subtracted $\langle \text{RM} \rangle$.
		\end{tablenotes}
	\end{threeparttable}
\end{table*}

\subsubsection{IC\,708}
\label{sec:ic708}

\begin{figure*}
	\centering
	\includegraphics[width=\linewidth]{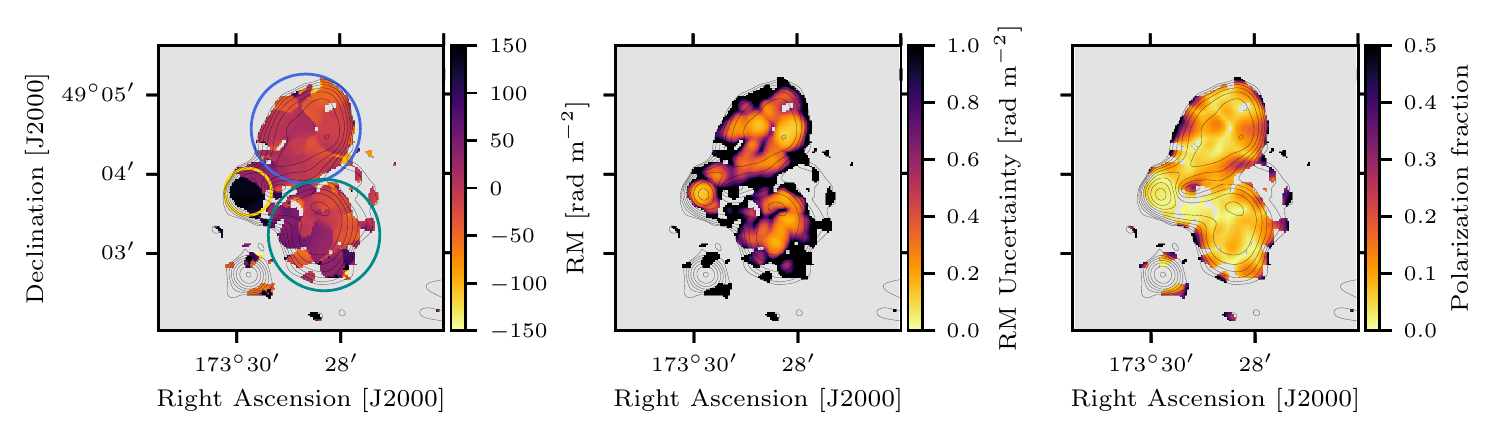}
	\caption{IC708. Contours show radio total intensity in increments of 1\,$\sigma$ from 5\,$\sigma$. Greyscale shows: (left) the observed peak rotation measure; (centre) the uncertainty on the observed rotation measure, $\sigma_{\phi}$; and (right) the polarisation fraction, $P/I$; all for the region where $P > 6\,\sigma_{\rm QU}$. The regions of the image are used to define the core, the north and south lobes of this galaxy. as described in Section~\ref{sec:ic708}, this is indicated by yellow, blue and cyan circles, respectively.}
	\label{fig:ic708}
\end{figure*}

The wide angle tail galaxy IC\,708 (Source~3; Table~\ref{tab:a1314-sources}) shows significant structure in polarisation across its lobes and this is reflected in the high RM dispersion value of $\sigma_{\rm RM} = 58.49$\,rad\,m$^{-2}$ for this source, see Figure~\ref{fig:ic708} and Table~\ref{tab:a1314_pol}. The host galaxy of IC\,708 has a much higher RM than the lobes: $\phi_{\rm peak} = 136.5 \pm 0.17$\,rad\,m$^{-2}$, suggesting that it is  affected by more local magneto-ionic structure from the interstellar medium of the host galaxy.
Table~\ref{tab:ic708_pol} shows RM statistics separately for the core region, and the North and South lobes of IC\,708. The areas used to define these regions are indicated in Figure~\ref{fig:ic708}. We note that the high RM dispersion indicated for the core region is heavily affected by the sharp transition in RM at the base of the jet for the Northern lobe.
From Table~\ref{tab:ic708_pol} it can also be seen that the North and South lobes have different signs for their average RM, indicating that there is a field reversal along the line of sight between these lobes.
\begin{figure*}
	\centering
	\subfloat[IC708]{
		\includegraphics[width=0.9\linewidth]{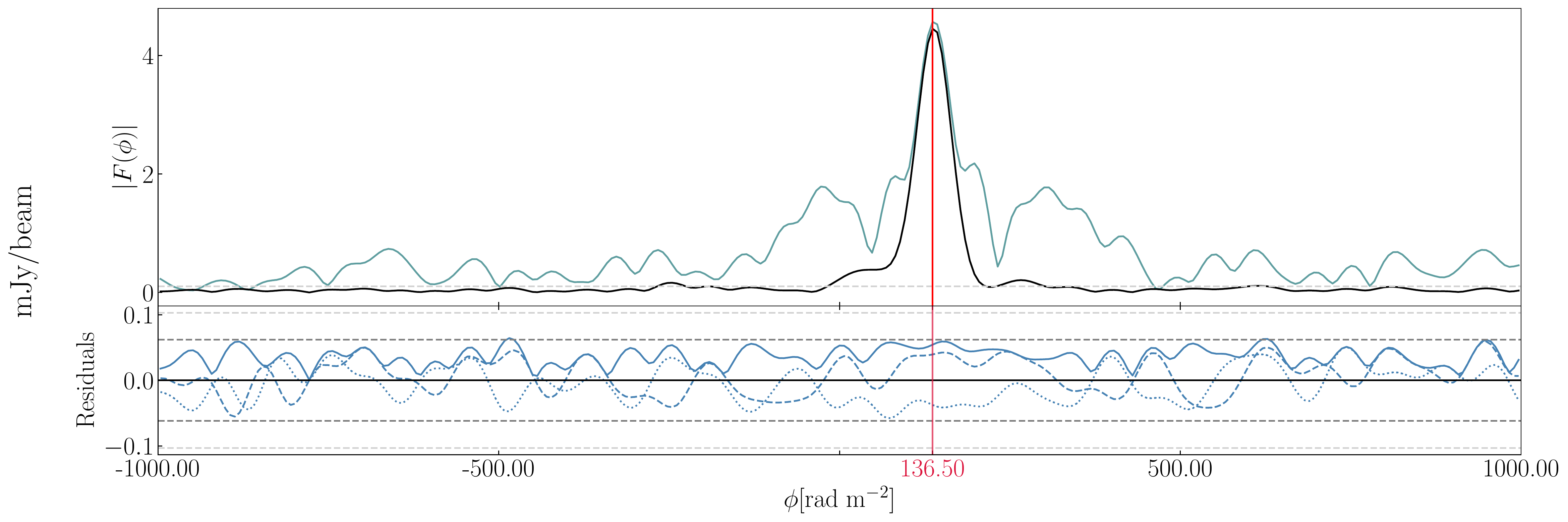}
		\label{fig:los-ic708}
	}

	\subfloat[IC711]{
		\includegraphics[width=0.9\textwidth]{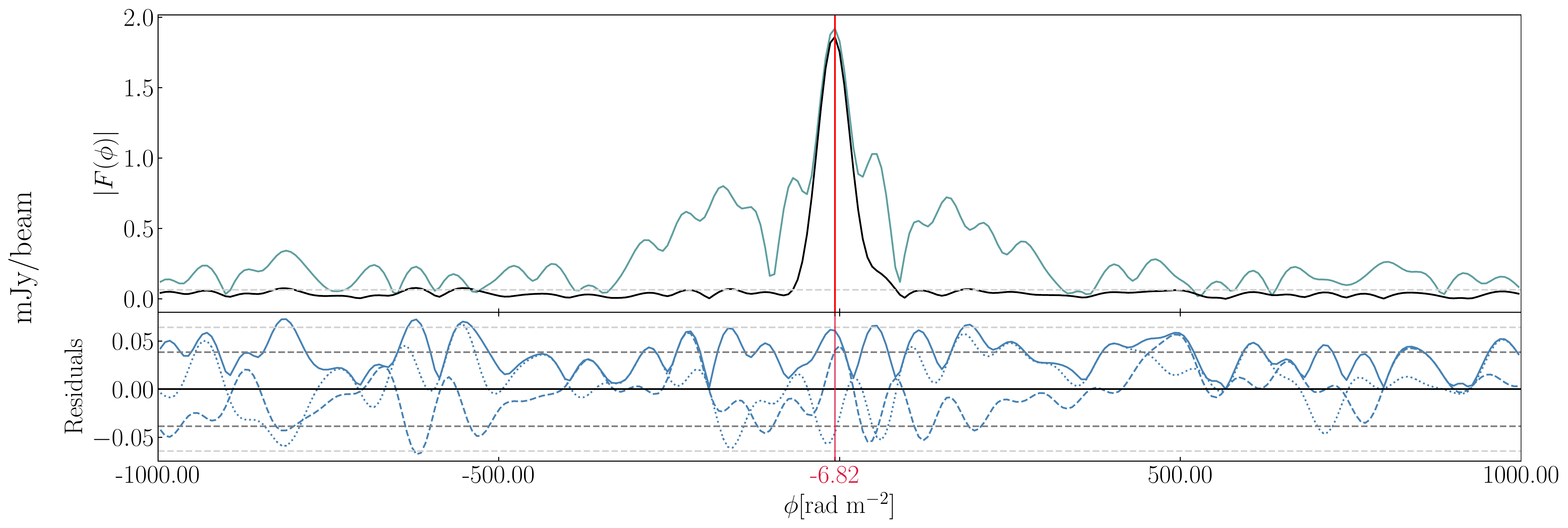}
		\label{fig:ic711_fd}
	}

	\subfloat[IC712]{
		\includegraphics[width=0.9\textwidth]{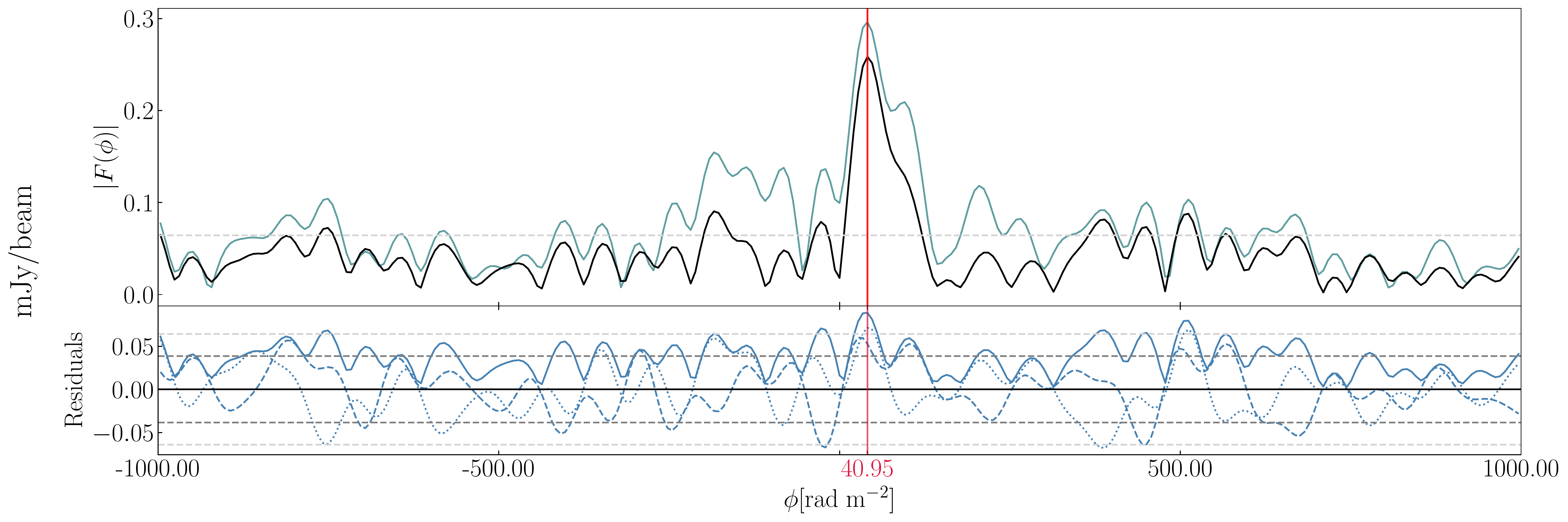}
		\label{fig:ic712_fd}
	}

	\caption{Lines of sight at the cores of IC708 (top), IC711 (middle) and IC712 (bottom). The upper panels in each figure show the dirty and the restored Faraday depths. The lower panels show the amplitude, the real and imaginary parts of the residual signal. Light and dark grey dashed lines show 5\,$\sigma$ and 3\,$\sigma$ boundaries, respectively.}
	\label{fig:los-ic-radgal}
\end{figure*}
%
%
%
\begin{table*}
	\caption{RM profile of polarized source IC\,708.}
	\label{tab:ic708_pol}
	\begin{threeparttable}[b]
		\centering
		\begin{tabular}{@{}ccccccccccc@{}}
			\hline
			Component  & z     & $\langle \text{RM} \rangle$ & $\langle \text{RM} \rangle$\tnote{a} & $\sigma_{\text{RM}}$ & med(RM)        & MAD(RM)        & med($\sigma_{\phi}$) & $n_{\text{beam}}$ \\
			&       & (rad m$^{-2}$)              & (rad m$^{-2}$)                       & (rad m$^{-2}$)       & (rad m$^{-2}$) & (rad m$^{-2}$) & (rad m$^{-2}$)       &                   \\\hline
			Core       & 0.032 & 88.33                       & 103.8                                & 50.9                 & 136.5          & 6.8            & 0.4                  & 5.0               \\
			North lobe & $-$   & -27.48                      & -11.3                                & 30.3                 & -6.8           & 13.7           & 0.5                  & 26.2              \\
			South lobe & $-$   & -5.96                       & 10.2                                 & 46.7                 & 6.8            & 27.3           & 0.7                  & 23.4              \\\hline
		\end{tabular}
		\begin{tablenotes}
			\item [a] Galactic-contribution-subtracted $\langle \text{RM} \rangle$.
		\end{tablenotes}
	\end{threeparttable}
\end{table*}

\subsubsection{IC\,711}
\label{sec:ic711}

The head-tail radio galaxy IC\,711 (Source~9; Table~\ref{tab:a1314-sources}) has a similarly high RM dispersion to IC\,708 with a value of $\sigma_{\rm RM} = 59.85$\,rad\,m$^{-2}$. The full extension of the long tail of emission associated with this source seen at longer wavelengths \citep{sebastian2017,wilber2019} is not visible in these higher frequency VLA data at this resolution. We note however that the discrete compact source listed as Source~6 in Table~\ref{tab:a1314-sources} and shown in Figure~\ref{fig:a1314_overlay}, is coincident with the location of the northern east-west extension of emission perpendicular to the main tail of IC\,711. This abrupt turn in the direction of the diffuse emission from IC\,711 was first noted by \cite{srivastava2020} and also detected by \cite{sebastian2017} and \cite{wilber2019}. Those works proposed that the turn in emission was potentially due to a disturbance caused by ram pressure or shocks propagating outwards in the intra-cluster medium. We suggest that alternatively this emission may not be part of the tail from IC\,711 but instead could be associated with Source~6.

\begin{figure*}
	\centering
	\includegraphics[width=\linewidth]{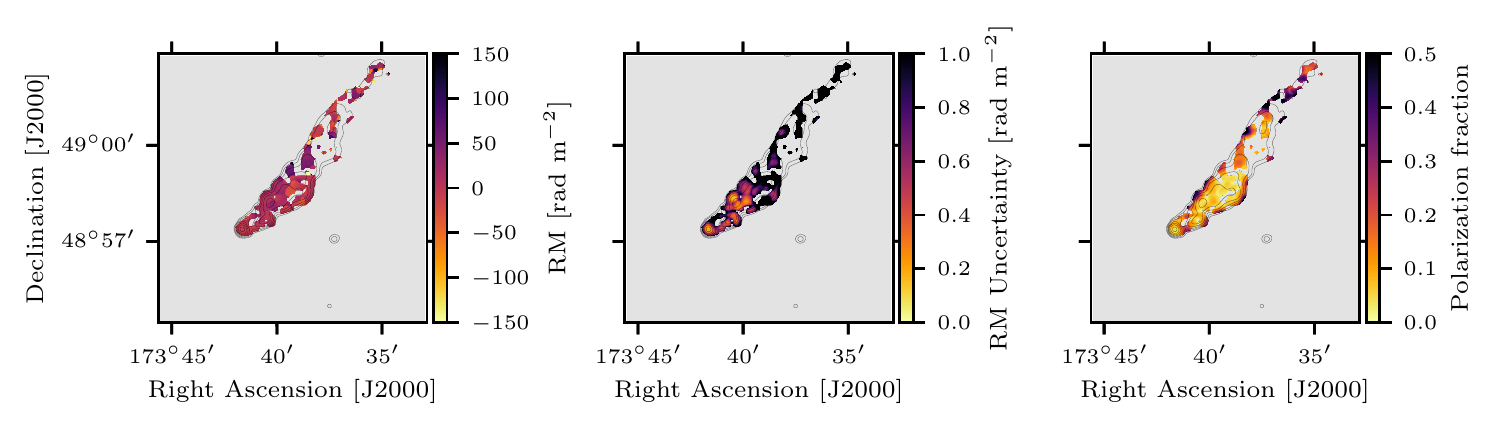}
	\caption{IC711. Contours show radio total intensity in increments of 1\,$\sigma$ from 5\,$\sigma$. Greyscale shows: (left) the observed peak rotation measure; (centre) the uncertainty on the observed rotation measure, $\sigma_{\phi}$; and (right) the polarisation fraction, $P/I$; all for the region where $P > 6\,\sigma_{\rm QU}$. }
	\label{fig:ic711}
\end{figure*}

\subsubsection{IC\,712}
\label{sec:ic712}

\begin{figure*}
	\centering
	\includegraphics[width=\linewidth]{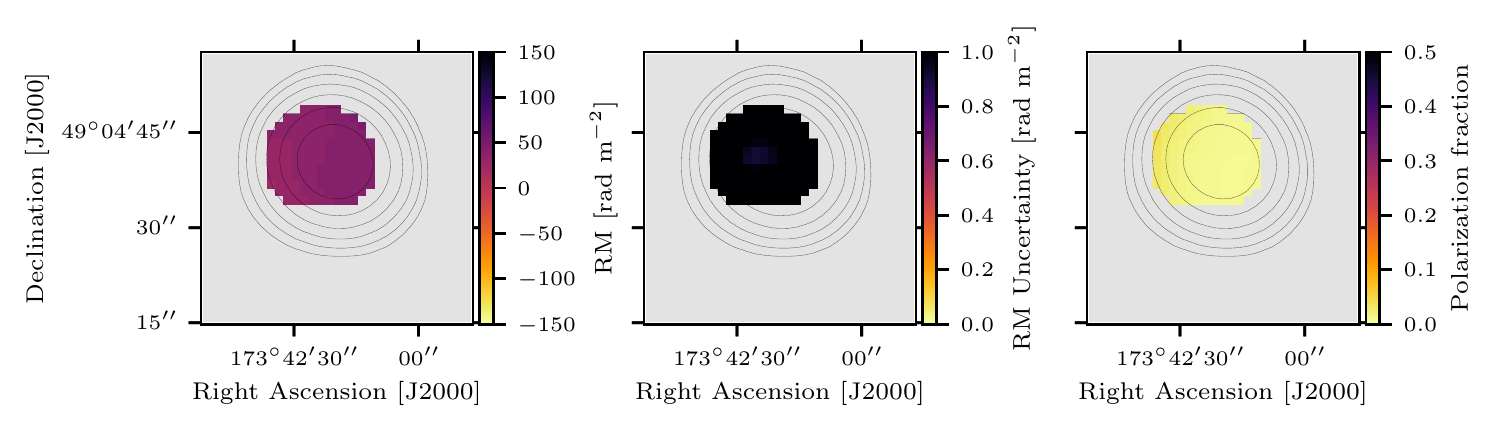}
	\caption{IC712. Contours show radio total intensity in increments of 1\,$\sigma$ from 5\,$\sigma$. Greyscale shows: (left) the observed peak rotation measure; (centre) the uncertainty on the observed rotation measure, $\sigma_{\phi}$; and (right) the polarisation fraction, $P/I$; all for the region where $P > 6\,\sigma_{\rm QU}$. }
	\label{fig:ic712}
\end{figure*}

IC\,712 (Source~10; Table~\ref{tab:a1314-sources}) is the brightest cluster galaxy in A1314 \citep{lin2004} and closest to the X-ray centre with a projected distance of $\sim 40$\,kpc, see Table~\ref{tab:a1314-sources}. 
The source appears compact in both Stokes~I and polarization, with a small offset of the polarized emission from the peak total intensity, see Figure~\ref{fig:ic712}.

The Faraday depth spectrum of IC\,712 shows a single Faraday thin structure, see Figure~\ref{fig:ic712_fd}, with a peak value of $\phi_{\rm peak} = 40.95\pm1.42$\,rad\,m$^{-2}$. It has been suggested that RMs smaller than the FWHM of the RMTF are expected to differ from their true value by about 5\% due to the presence of leakage at $\phi=0$\,rad\,m$^{-2}$ \citep{jagannathan2017}, which would increase the uncertainty on this measurement to $\Delta\phi_{\rm peak} \simeq 3$\,rad\,m$^{-2}$.

\subsection{RM profiles}

\begin{figure*}
	\centering
	\includegraphics[width=\textwidth]{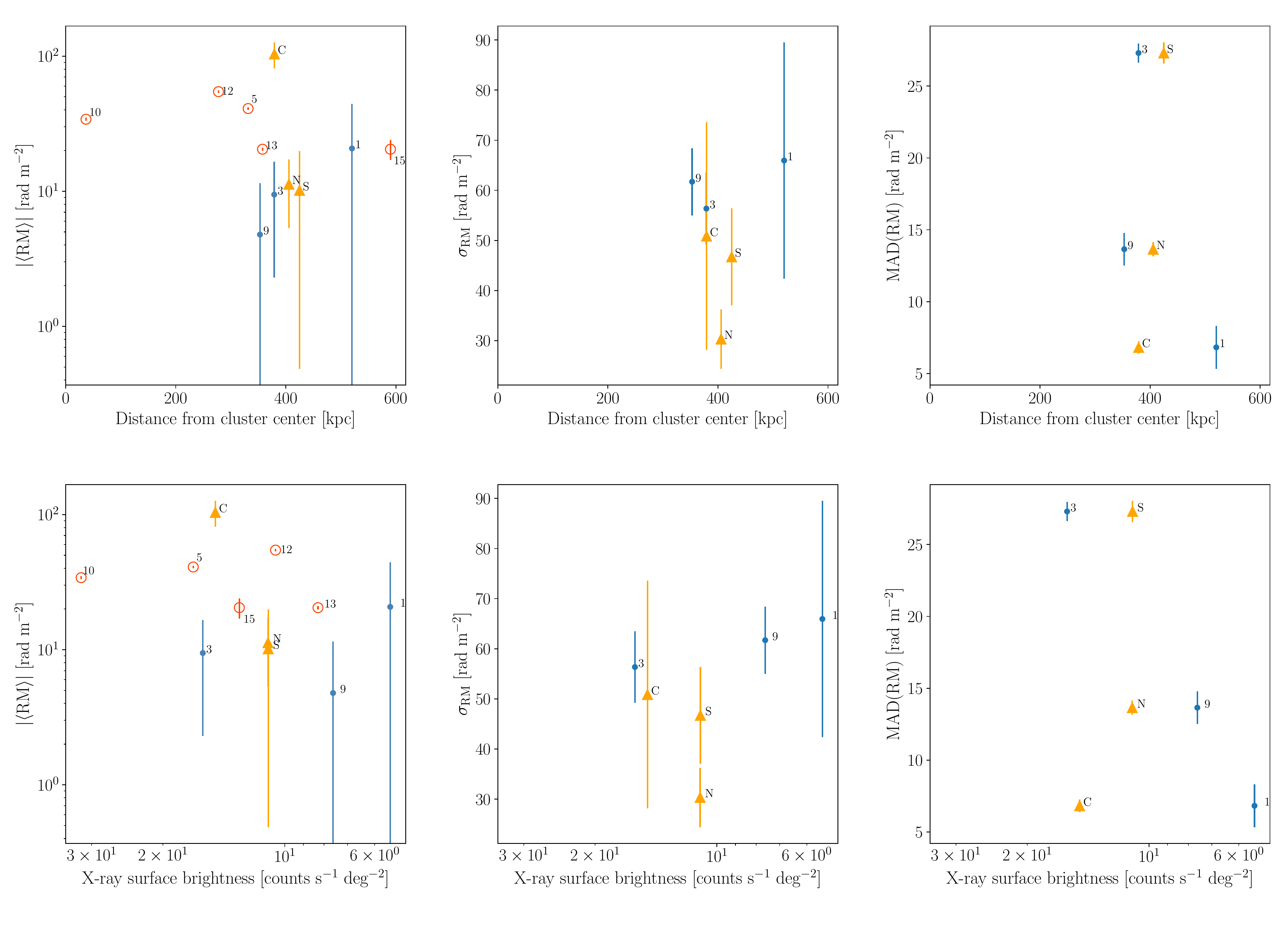}
	\caption{$|\langle \text{RM} \rangle|$, $\sigma_{\text{RM}}$ and MAD(RM) of the polarized sources in the Abell 1314 cluster plotted against the projected distance of each source to the X-ray center (top row) and against the X-ray surface brightness at the position of peak polarization for each source (bottom row). The uncertainties of the $|\langle \text{RM} \rangle|$ and $\sigma_{\text{RM}}$ are the $\pm 1\sigma$ considering $n_{\text{beam}}$ samples for each source. Sources that cover an area with fewer than 5\,$n_{\text{beam}}$ are shown as open red circles and are not considered in the $\sigma_{\text{RM}}$ and MAD(RM) plots. Uncertainties for MAD(RM) are derived from the median error on the RM measurement, $\text{med}(\sigma_\phi)$. The detected polarized sources are numbered according to Table \ref{tab:a1314_pol}. The components of IC708 are shown with yellow triangles and named as C, N and S for core, north lobe and south lobe, respectively, see Table~\ref{tab:ic708_pol}.}
	\label{fig:rm_profiles}
\end{figure*}

Radial profiles of the absolute average RM, RM dispersion and median absolute deviation for all sources detected in polarization are shown in Figure~\ref{fig:rm_profiles}. The radial distance of each source is calculated as the projected distance from the X-ray peak (11h34m51.36s +49d05m27.6s) to the position of peak polarisation within the given source. Sources that extend over areas smaller than $n_{\rm beam}=5$ are excluded from the $\sigma_{\rm RM}$ and MAD(RM) profiles, as they have insufficient independent samples to be considered. The separate components of radio galaxy IC\,708 are also indicated in Figure~\ref{fig:rm_profiles}.

Unlike the work of \cite{stuardi2021}, which performed a similar analysis for the galaxy cluster Abell\,2345, a radial trend is seen only in the absolute average RM, with RM dispersion and MAD(RM) values not indicating any clear systematic behaviour. These results suggest that for A1314, the local RM contribution to each source dominates over the contribution of the ICM. This is particularly notable for the galaxy IC\,708 for two reasons: firstly the average RM of the host deviates substantially from the radial behaviour seen in the absolute average RM for other sources, indicating that the host galaxy of IC\,708 has a significant local contribution to its RM; secondly, the difference in the MAD values between the North and South lobes of IC\,708 indicates a variance that also cannot be accounted for by the changes in the local ICM, which is quite similar for these closely located regions. This second point is supported by the lower panels of Figure~\ref{fig:rm_profiles}, which show the RM profiles as a function of X-ray surface brightness, a proxy for electron density in the ICM. Once again, although the surface brightness local to both lobes of IC\,708 is similar, the MAD(RM) values are highly discrepant. As noted by \cite{stuardi2021}, the MAD(RM) is expected to be the most robust estimator of RM dispersion and hence the best measure with which to distinguish local from large-scale environments.

Furthermore, the Faraday depth spectrum towards the core of IC\,708, see Figure~\ref{fig:los-ic708}, shows a potential deviation from the Faraday-simple structure that would be expected from an external Faraday-thin screen such as the ICM. The Faraday depth spectrum for the host of IC\,708 shows an extension of polarised emission from the main peak that is detected at a significance of $\sim20\,\sigma'_{\phi}$. This structure could be caused by the Faraday thin peak being embedded in a more local Faraday thick region. Such structure would be geometrically consistent with the presence of a compact source embedded within an envelope of emitting and rotating material. Or alternatively, rather than mixing of Faraday rotation and emission, such complexity could be caused by structure in an unresolved screen associated with the host galaxy. Slight deviations from purely  Faraday thin structure are also seen in the Faraday depth spectra for IC\,711 and IC\,712, see Figures~\ref{fig:ic711_fd}~\&~\ref{fig:ic712_fd}; however, we note that these sources are not detected at such high significance in polarisation as IC\,708 and therefore the presence of additional structure is less well-defined.

Unlike IC\,708, the head-tail radio galaxy IC\,711 has an average RM that is low compared to the general radial decrease observed for other sources. The reason for this discrepancy can be seen in Figure~\ref{fig:a1314_overlay}, where it is clear that the majority of the radio emission recovered for IC\,711 lies outside the main X-ray emitting region and consequently this source will experience lower rotation than sources embedded within the higher density regions of the ICM. This is also clear from the lower panel of Figure~\ref{fig:rm_profiles}, where IC\,711 (Source 9) is associated with the second lowest surface brightness measurement.

\subsection{Discussion}
\label{sec:discussion}

The lack of a clear radial decrease in RM dispersion or MAD(RM) for Abell\,1314, as well as a lack of equivalent behaviour with electron density, suggests that the Faraday rotation exhibited by the galaxies associated with this cluster is dominated by their local environments, rather than the surrounding intra-cluster medium. One possibility to explain this circumstance is that all of the galaxies with extended structure are in the foreground of this cluster, but this seems unlikely given the similarity in redshifts. Alternatively, the comparatively low mass and consequently poor environment of Abell\,1314 ($M_{500} = 0.68\times 10^{14}$\,M$_{\odot}$) compared to more massive and richer systems such as Abell\,2345 ($M_{500} = 5.91\times 10^{14}$\,M$_{\odot}$) may have resulted in a system where the cluster medium plays a sub-dominant role in terms of Faraday structure, compared to the local environments of its more powerful constituent galaxies.

This interpretation is supported by the Faraday depth spectra of the radio galaxy IC\,708, which exhibits more complex Faraday structure than the equivalent galaxies in Abell\,2345, which show no additional structure above a significance of 6\,$\sigma_{\phi}$. The detail of such Faraday structure is therefore potentially important for understanding which cluster galaxies can be used to constrain models for cluster magnetic fields, i.e. which galaxies are likely to have structure dominated by the ICM rather than their local environment.

\section{Conclusions}
\label{sec:conclusion}

In this work we have introduced the object-oriented {\tt cs-romer} framework for reconstructing Faraday depth structure from radio polarimetric data using a compressed sensing formalism that regularises the under-constrained RM-CLEAN optimisation problem. {\tt cs-romer} improves upon previous applications in a number of ways. Firstly by simulating sources directly in frequency (data) space, and using the NUFFT in order to calculate the values of irregular $\lambda^2$-space measurements from regularly spaced Faraday depth when reconstructing. Secondly, the framework presents a variety of filters for discrete and undecimated wavelet transforms and an option to combine them with a delta function basis. Additionally, {\tt cs-romer} not only uses a prior that imposes sparsity on Faraday depth space or wavelet space, but it also provides convex regularisation functions such as TV and TSV to smooth the reconstructed signal. These features make the framework more generalisable than if just imposing sparsity on the coefficients using the $L_1$-norm.

The framework also implements optional astrophysical features such as derotation of Galactic Faraday rotation, which can be applied directly in $\lambda^2$-space as a user-selected pre-processing step based on external measures of the Galactic foreground. Additionally, at the user's discretion, the framework also adds the option to correct for the spectral dependency of the radio polarisation data. This can be done using either a direction-independent spectral index or a direction-dependent spectral index image resulting from an MFS deconvolution algorithm such as {\sc casa}'s {\tt tclean}.

Using simulated radio data for the VLA telescope, sampled directly in frequency rather than $\lambda^2$-space, we have demonstrated the {\tt cs-romer} framework for different Faraday structure scenarios and under different observational conditions, including a wide range of signal-to-noise and RFI flagging instances. By comparing several different evaluation metrics, we have used these simulated scenarios in order to select the wavelet basis that best represents the Faraday depth data for a given observational set-up. Even though in this work the preferred wavelet is the delta function basis, a wider range of simulations for different telescopes and Faraday structures may identify cases where other bases are preferred. Similar simulations may be used to represent the data from a wide range of radio telescopes with alternative spectral resolutions and frequency coverages in order to determine the appropriate basis function(s) for other datasets.

We demonstrate the {\tt cs-romer} framework on real data from the JVLA telescope towards the low-mass galaxy cluster Abell\,1314, using the optimal basis function determined through simulations and incorporating de-rotation of the Galactic foreground as well as direction-dependent spectral corrections. The results from this analysis indicate  that individual galaxies within Abell\,1314 show Faraday depth behaviour that deviates from what is expected for a Faraday-thin screen, suggesting that the Faraday rotation exhibited by the galaxies in this cluster, most notably IC\,708, is dominated by local magneto-ionic structures rather than the large-scale intra-cluster medium.

Potential future improvements to the {\tt cs-romer} framework might include the implementation of a more flexible optimization algorithm, such as SDMM \citep{Moolekamp2018}, which would allow the framework to generalize the compressed sensing problem further by adding several constraint functions or regularizations. Additionally, although {\tt cs-romer} currently uses the {\tt pywavelets} package in order to provide more than a hundred discrete wavelets, the software does not yet include any continuous wavelets. The implementation of continuous wavelets would make {\tt cs-romer} more flexible as a user could program their own wavelets in order to decompose and reconstruct signals. Furthermore, although we have currently adopted the error bound calculation from \cite{2014MNRAS.439.3591C, purify2}, the study of multiple regularization parameters remains an open topic of research. There have been many studies about regularization parameter selection criteria \citep[see e.g.][]{KARL2005183, Hansen00thel-curve, SHI201872} and as a future addition we anticipate that implementing an L-hypersurface criteria using the fixed-point optimization method from \cite{Belge_2002} would be advantageous. For all of these potential extensions to the existing framework, we note that current and future users can take advantage of {\tt cs-romer}'s object-oriented programming paradigm in order to facilitate the inclusion of enhancements in a straight forward manner. We encourage the reader to clone or fork {\tt cs-romer} on Github at \url{https://www.github.com/miguelcarcamov/csromer}.

\section*{Acknowledgements}

We thank the referee for making very constructive comments that improved the manuscript.

MC acknowledges support from the National Agency for Research and Development (ANID) PFCHA/DOCTORADO BECAS CHILE/2018-72190574. AMS \& ELA gratefully acknowledge support from the UK Alan Turing Institute under grant reference EP/V030302/1. ELA gratefully acknowledges support from the UK Science \& Technology Facilities Council (STFC) under grant reference ST/P000649/1.

Analysis in this paper uses observations obtained with \emph{XMM-Newton}, an ESA science mission with instruments and contributions directly funded by ESA Member States and NASA, and the Karl G. Jansky Very Large Array (VLA), operated by the National Radio Astronomy Observatory (NRAO). NRAO is a facility of the National Science Foundation operated under cooperative agreement by Associated Universities, Inc.

\section*{Data Availability}

This work uses A1314 archival data from the \emph{XMM-Newton} (ObsID: 0149900201) and VLA data found in the archive under priject ID 18A-172. Both datasets are publically available at \url{http://nxsa.esac.esa.int/nxsa-web/} and \url{https://archive.nrao.edu/archive/advquery.jsp}, respectively. Full code for the {\tt cs-romer} framework is available on Github at \url{https://www.github.com/miguelcarcamov/csromer}, including example notebooks for analysing both the simulated and real radio data presented in this work.



\bibliographystyle{mnras}
\bibliography{a1314} 


\appendix
\onecolumn

\section{Data corruption evaluation}
\label{app:datacorruption}

\begin{figure*}
	\centerline{\includegraphics[width=\linewidth]{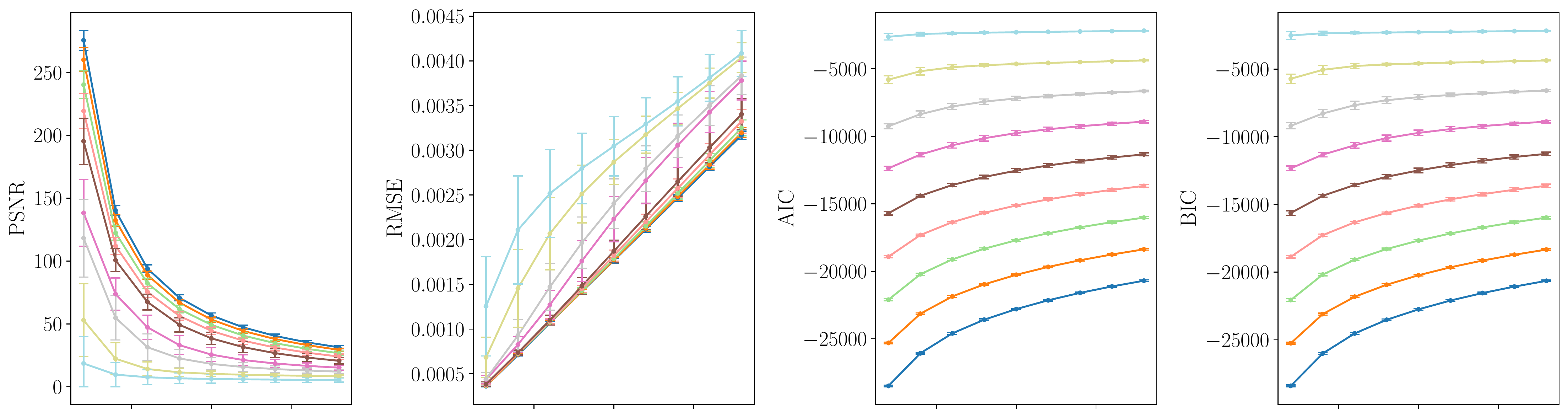}}
	\centerline{\includegraphics[width=\linewidth]{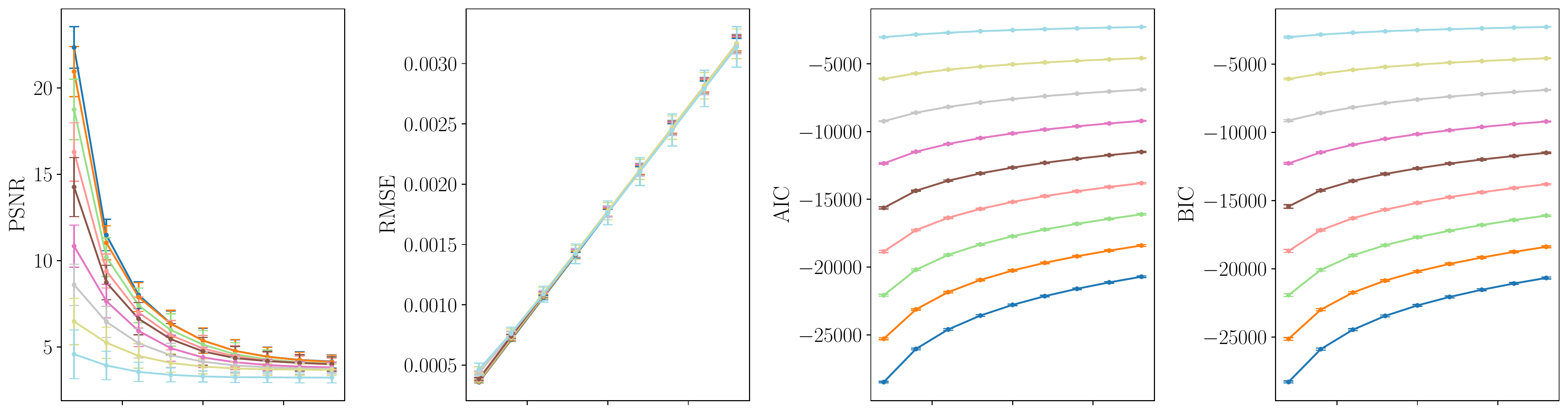}}
	\centerline{\includegraphics[width=\linewidth]{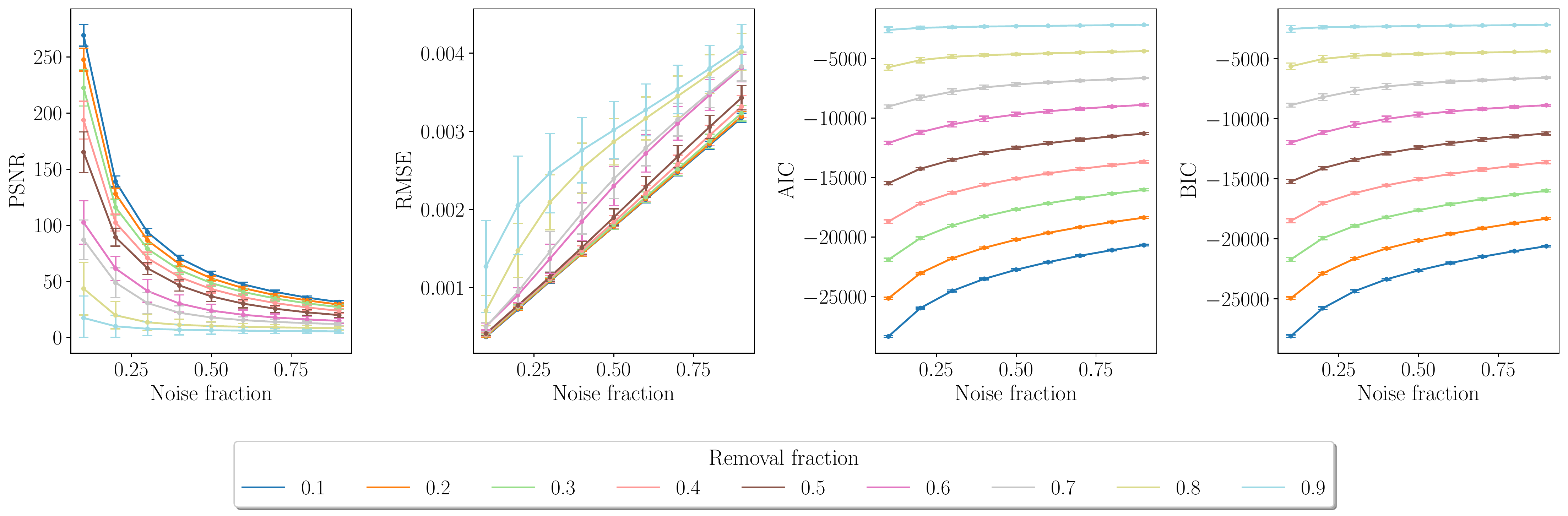}}
	\caption{PSNR, RMSE, AIC and BIC for the delta function basis reconstruction. First row shows scenario 1, second row for scenario 2 and third row for scenario 3.}
	\label{fig:deltafbasis_aicbic}
\end{figure*}

\begin{figure*}
	\centerline{\includegraphics[width=\linewidth]{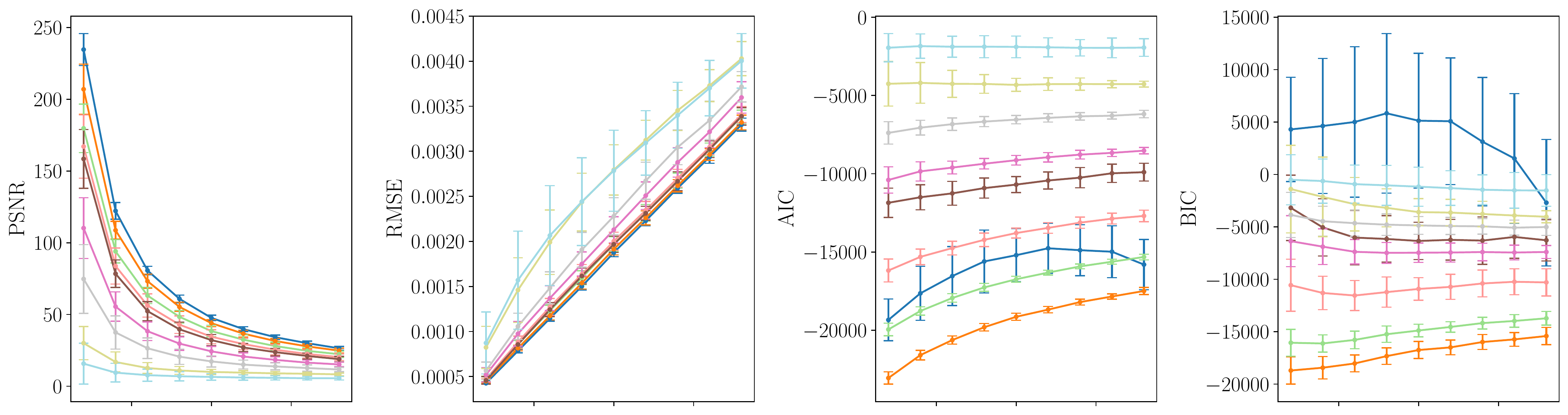}}
	\centerline{\includegraphics[width=\linewidth]{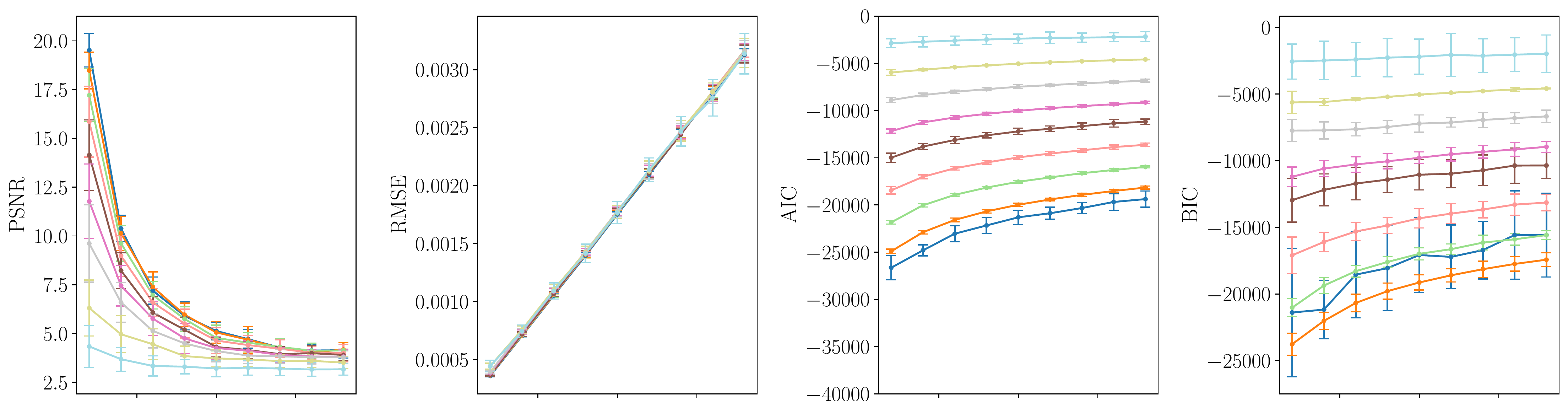}}
	\centerline{\includegraphics[width=\linewidth]{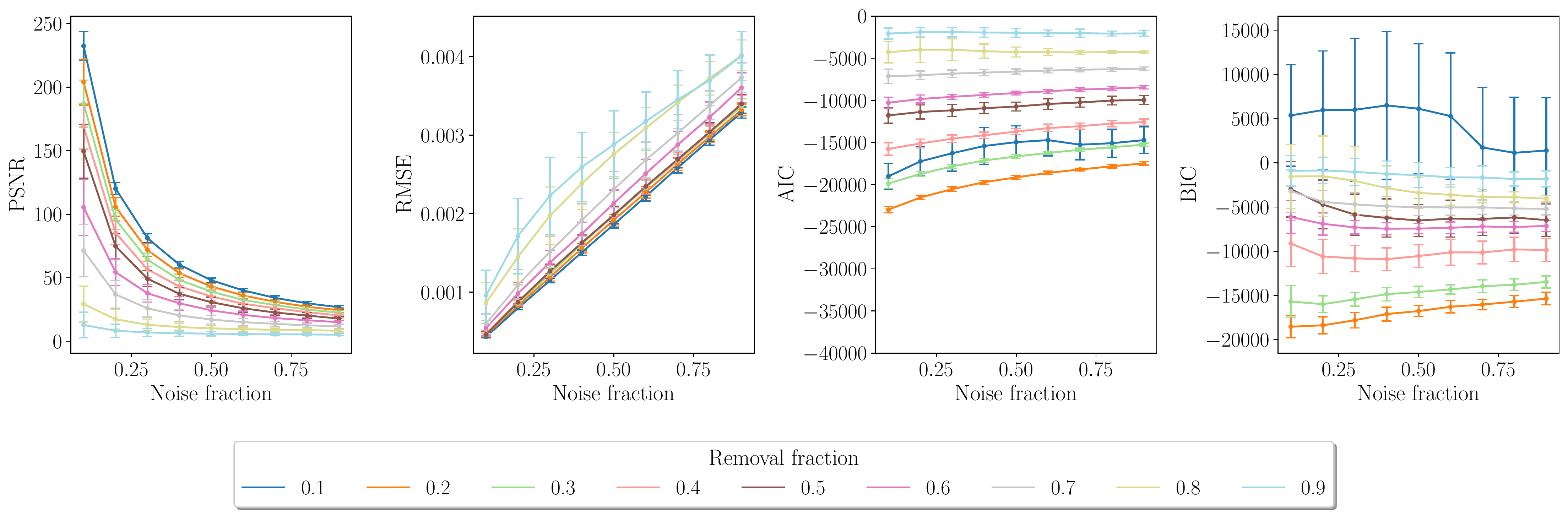}}
	\caption{PSNR, RMSE, AIC and BIC using an undecimated wavelet basis reconstruction. First row shows scenario 1, second row for scenario 2 and third row for scenario 3. We have used Haar wavelets for all scenarios.}
	\label{fig:undecimated_csbasis_aicbic}
\end{figure*}

Figures~\ref{fig:deltafbasis_aicbic}~$-$~\ref{fig:undecimated_csbasis_aicbic} show the results of 50 realisations for the three scenarios using removal fractions and noise fractions from 0.1 to 0.9. Figure~\ref{fig:deltafbasis_aicbic} show the results using the delta function basis and Figure \ref{fig:undecimated_csbasis_aicbic} show the results using the undecimated wavelet transform (UWT). For more details of the results of these experiments see Section~\ref{sec:corruption}.

\section{XMM Analysis}
\label{app:xray-xmm-analysis}

\subsection{Data reduction}
\label{app:xrayreduction}

A1314 was observed by XMM-Newton in November 2003 during rev. 725 (ObsID: 0149900201) with a total exposure time of 18.4\,ks. The observation was performed in full frame mode for the MOS cameras and the PN detector, all using the medium filter. Observation data files (ODFs) were downloaded from the XMM-Newton archive and processed with the XMMSASv19.1.0 software for data reduction (Gabriel et al. 2004). Processing was performed following the steps of the XMM-Newton Extended Source Analysis Software (XMM-ESAS) pipeline (Snowden et al. 2008; Kuntz \& Snowden 2008). We used the tasks {\tt emchain} and {\tt epchain} to generate calibrated event files from raw data. We excluded all the events with $\rm PATTERN>4$ for PN data and with $\rm PATTERN>12$ for MOS data. Bright pixels and hot columns were removed by applying the expression $\rm FLAG==0$. The data were observed to be free of significant periods of high background induced by solar flares and the statistical thresholding provided by the {\tt mos-filter} and {\tt pn-filter} tasks was used to remove any residual contamination. The remaining exposure times after cleaning were 16.7\,ks for MOS1, 16.9\,ks for MOS2, and 12.9\,ks for PN.

Point-like sources were detected using the XMM-ESAS task {\tt cheese}, which combines data from all three EPIC detectors to provide source lists and masks for removing such sources from images and spectra. The compact X-ray source associated with the host galaxy of IC708 was masked manually as it lies close to CCD boundaries for the MOS1 and MOS2 detectors. The {\tt mos-spectra} and {\tt pn-spectra} tasks were used to create redistribution matrix files (RMFs), auxiliary response files (ARFs) and exposure maps for the full field of view of each detector, and these were then used to create models of the quiescent particle background (QPB) in each case. The rate-hardness plots for each of the CCDs with unexposed corners were inspected to check for CCDs in anomalous states; no CCDs were excluded on this basis. A combined $0.3 - 2.0$\,keV data set was created assuming a power-law spectrum with $\alpha = 0.7$ and a $2\times 10^{20}$\,cm$^{-2}$ {\sc Hi} absorption. The resulting $0.3 - 2.0$\,keV combined background-subtracted exposure-corrected image with point sources masked is shown in Figure~\ref{fig:annuli}.

\subsection{X-ray Spectra}
\label{sec:spectra}

X-ray spectra were fitted jointly to data from the three EPIC detectors using the Xspec software \citep{xspec}. The fitted model included instrumental spectral lines at energies of 1.496\,keV (Al Kalpha) and 1.75\,keV (Si Kalpha) in the MOS data and at energies of 1.496\,keV (Al Kalpha) and near 8\,keV (Cu; 71, 7.5, 7.9, 8.2, 8.5\,keV) in the PN data, spectral lines due to solar wind charge exchange at 0.56 and 0.65\,keV, and a cosmic background component linked for all spectra and constrained additionally through the inclusion of a background spectrum extracted from an annulus between $1 - 2^{\circ}$ radius from the X-ray peak of A1314 using the ROSAT All-Sky Survey diffuse background maps\footnote{\url{https://heasarc.gsfc.nasa.gov/cgi-bin/Tools/xraybg/xraybg.pl}} \citep{rassbkg} and associated response files. The cosmic background component is modelled as a combination of a cool ($E\sim0.1$\,keV) unabsorbed thermal component representing emission from the Local Hot Bubble or heliosphere, a cool ($E\sim0.1$\,keV) absorbed thermal component representing emission from the cooler halo, a higher temperature ($E\sim0.25-0.7$\,keV) absorbed thermal component representing emission from the hotter halo and/or intergalactic medium, and an absorbed power law with $\alpha\sim1.46$ representing the unresolved background of cosmological sources. The cluster emission itself is modelled using an absorbed thermal component, represented by an apec model.

In addition to the above components, broken power-laws are included in the fits for all three of the EPIC detectors to account for any residual soft-proton background emission that was not been completely removed by the time-dependent filtering applied during the data reduction process.

The non-detector components of the fit were scaled by solid angle separately for each detector using values from the XMMSAS {\tt proton\_scale} task. Absorption was fixed using the Galactic value\footnote{\url{https://heasarc.gsfc.nasa.gov/cgi-bin/Tools/w3nh/w3nh.pl}} of $N_{\rm H} = 1.53\times 10^{20}$\,cm$^{-2}$ \citep{hi4pi} and a redshift value of $z=0.034$ was used for all fits.

\subsection{Radial temperature profile}
\label{sec:temp}

\begin{figure}
	\centering
	\includegraphics[width=0.45\textwidth]{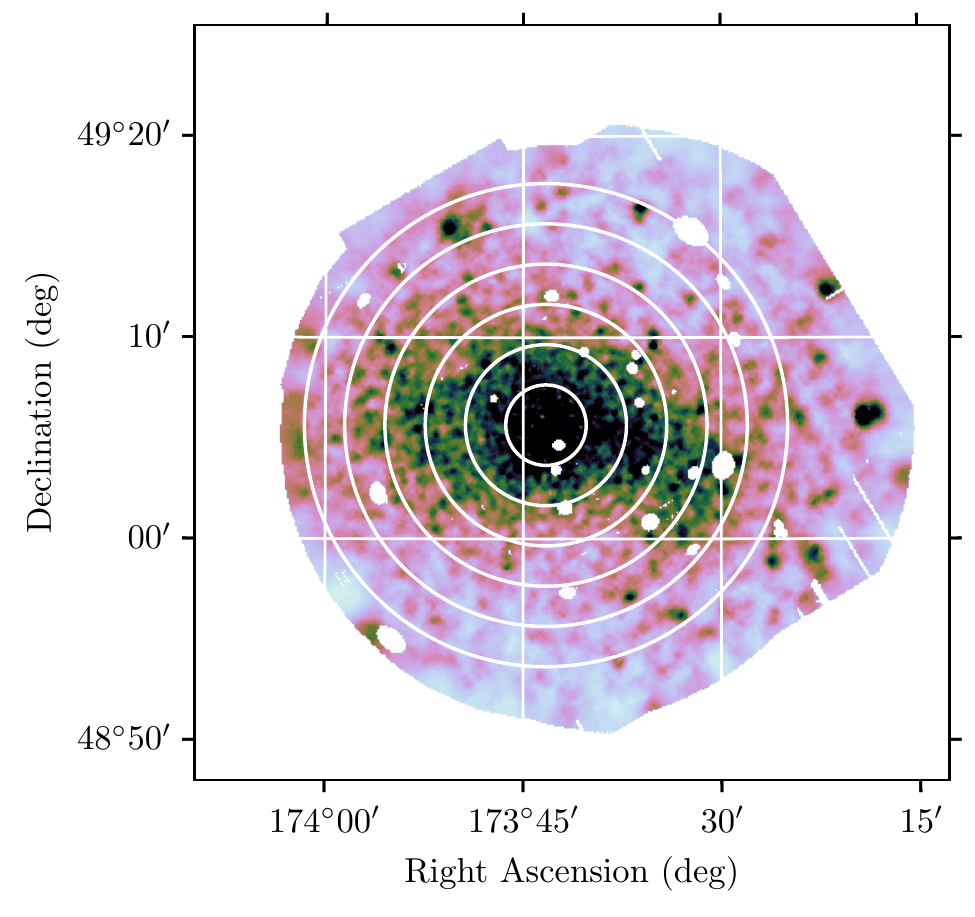}
	\caption{Regions used for extracting radial cluster properties are overlaid on the $0.3-2.0$\,keV adaptively-smoothed, background-subtracted and exposure-corrected image from the combined MOS1, MOS2 \& PN exposures. The data have point sources masked as described in Section~\ref{app:xrayreduction}.}
	\label{fig:annuli}
\end{figure}

\begin{figure}
	\centering
	\includegraphics[width=0.425\textwidth]{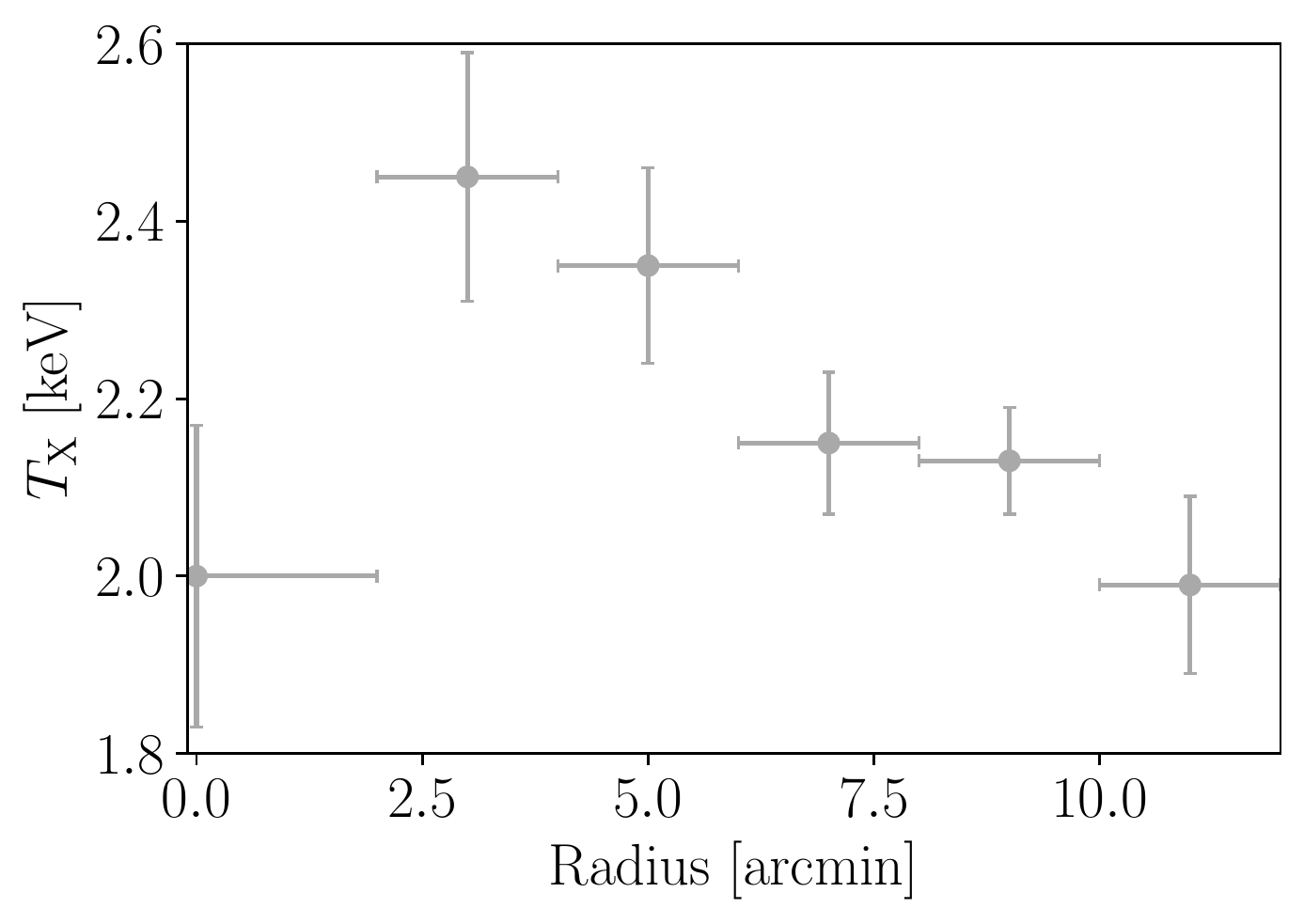}
	\caption{Radial temperature profile for Abell\,1314. Data points correspond to spectral fits of the emission in the core and annular regions shown in Figure~\ref{fig:annuli}}.
	\label{fig:temp}
\end{figure}

A radial temperature profile was extracted from the data by fitting spectra from five annular regions with a width of two arcminutes each, as well as a central core region with a radius of two arcminutes. These regions are shown in Figure~\ref{fig:annuli}. Spectra were fitted independently for each region as described in Section~\ref{sec:spectra}. The results from these fits are shown in Figure~\ref{fig:temp}, where it can be seen that the central region is marginally cooler than the outer regions of the cluster.
%

\bsp	
\label{lastpage}
\end{document}